\documentclass{tMAM2e}

\usepackage{epstopdf}
\usepackage[bookmarksnumbered,colorlinks,bookmarks,citecolor=blue,linkcolor=blue,urlcolor=blue,breaklinks,linktocpage]{hyperref}
\usepackage[all]{hypcap}
\usepackage{subfigure}

\theoremstyle{plain}

\theoremstyle{definition}

\theoremstyle{remark}

\begin{document}



\title{Well Temperaments based on the Werckmeister definition\\From Werckmeister, passing Vallotti, Bach and Kirnberger,\\ to Equal Temperament.\\Inspired by Kelletat and Amiot}

\author{
\name{Johan Broekaert$^{\ast}$\thanks{$^\ast$Corresponding author. Email: broekaert.devriendt@gmail.com}
}
\affil{Faculty of Engineering Sciences, Catholic University of Leuven, Leuven,  Belgium}
\received{}
}

\maketitle

\begin{abstract}
Mathematical steps are developed to obtain optimised well tempered model ``temperaments'', based on
 a musical Well Temperament definition by H. Kelletat, that is derived from A. Werckmeister. This supports
objective mathematical ranking of historical temperaments. Historical
temperaments that fit best with mathematical models are often installed on organs. Musical appreciation
and mathematical ranking of historical temperaments are in agreement.
The diatonic C-major impurity of Well Temperaments lies between that of an
elaborated optimal model and that of Equal Temperament. Historical Well
Temperaments join these limits very closely and the list is amply filled in small
rising steps. Hence, it achieves little to develop new Well Temperaments.\\
The developed RMS computing modules can also be used for precise and accurate
recognition of historical temperaments.\\

\noindent \textbf{Online Supplement:} Spreadsheet with calculations and solutions

\end{abstract}





\noindent ``Ce qui se conçoit bien s'énonce clairement, et les mots pour le dire arrivent aisément.''\footnote{What is well conceived can be stated clearly, and words flow with ease.}
\begin{flushright}
\citep{Boileau1674}
\end{flushright}
%
%
\section{Issue}
Ever since Antiquity, temperaments are a subject of discussion. Well Temperaments 
(WT)\footnote{Nowadays WT are also often named ``Circulating Temperaments''} emerged in the 17-th century.
Many differing WT are applied, but none is universally accepted. Musical WT criteria are
pretty hard to translate into comprehensive objective evaluations or classification \citep[][pp. 274-290]{Hall1973}. 
All this results in the fact that the choice of a musical temperament
remains part of the artistic freedom of musicians.

A novel and objective WT evaluation is proposed, by comparing historical WT with
mathematically optimised WT models, with minimum diatonic impurity for C-major; the
impurity being calculated according to\footnote{RMS: square Root of the Mean of the Squares}\\
%
\begin{equation}
RMS_{Diatonic\ impurity}=\sqrt{\frac{\sum_{Weighted} (\mathrm{C}\mathrm{-}\mathrm{major}\ diatonic\ thirds\ \mathrm{and}\ fifths\ impurities)^2}{\sum(weights\ of\ intervals)}}
\label{equ 1}
\end{equation}\\
followed by a similar optimisation for fifths outside the diatonic C-major.
%
%
\subsection{Well Temperaments}
\begin{figure} [h]
\begin{center}
\includegraphics{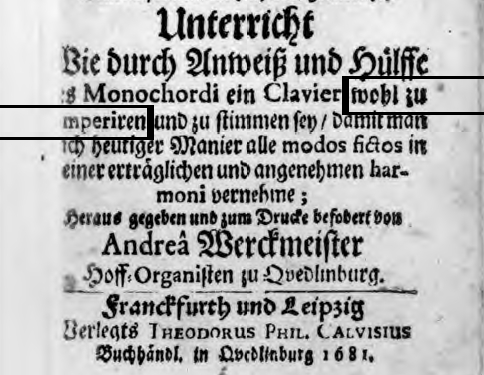}
\caption{Orgelprobe 1681}
\label{fig. 1}
\end{center}
\end{figure}

The German term for Well Temperament is
``Wohltemperiert''. First Wohltemperiert use in
writing was probably by Werckmeister A. \citeyearpar[][title page]{Werckmeister1681}, see fig.\ref{fig. 1} \citep[][p. 18]{Norback2002},
 and was also used later \citeyearpar{Werckmeister1686, Werckmeister1689, Werckmeister1698}.

%
\noindent
Werckmeister's WT requirements are still
accepted and referred to nowadays \citep[][pp. 25-26]{Norback2002}. The Werckmeister III
temperament became widespread and famous. The ``Wohltemperiert'' term probably
became popular, for it was used in the title of ``Das Wohltemperirte Clavier'' (Bach J. S.
1722). It should however be noticed, that the musical education of J. S. Bach was based on
meantone \citep[][p. 21, lines 3 to 9]{Kelletat1981}. F. Marpurg \citeyearpar[][pp. 210-223; §§ 227-237]{Marpurg1776} published an opinion,
rejected by J. Kirnberger, -a student and follower of J. S. Bach- \citep[][p. 42]{Kelletat1981}, that J. S. Bach used the equal temperament.
This opinion became widely published, copied, and accepted. The assumed WT and 12TET
equivalence and 12TET practice by J. S. Bach, was only rejected in modern times; probably first by B. Bosanquet \citeyearpar[][pp. 29-30]{Bosanquet1876}, 
and only much later, again but thoroughly, by H. Kelletat \citeyearpar{ Kelletat1957,  Kelletat1960}, which led to a
breakthrough. Werckmeister’s publications have lead to a musical WT definition
elaborated by H. Kelletat \citeyearpar[][page 9, first paragraph]{ Kelletat1981}\footnote{Wohltemperierung heißt mathematisch-akustische und praktisch-musikalischen Einrichtung
von Tonmaterial innerhalb der zwölfstufigen Oktavskala zum einwandfreien Gebrauch in
allen Tonarten auf der Grundlage des natürlichharmonischen Systems mit Bestreben
möglichster Reinerhaltung der diatonische Intervalle. \\
Sie tritt auf als proportionsgebundene, sparsam temperierende Lockerung und Dehnung des
mitteltönigen Systems, als ungleichschwebende Semitonik und als gleichschwebende Temperatur.}:
\begin{quote}
WT means a mathematical-acoustic and musical-practical organisation of the tone
system within the twelve steps of an octave, so that impeccable performance in all
tonalities is enabled, based on the extended just intonation (natural-harmonic tone
system), while striving to keep the diatonic intervals as pure as possible.

\noindent While tied to given pitch ratios, this temperament occurs as a thriftily tempered
smoothing and extension of the meantone, as unequally beating semitonics and as
equal (equally beating) temperament.
\end{quote}
Later on many assumptions are made and published, about what has come to
be called ``Bach temperaments''; mathematical proposals arise also, about how to assess or
create WT \citep{Kellner1977, Goldstein1977, Sankey1997, Polansky2008,
Amiot2010, Farup2013, Liern2015}. This paper is mainly in line with Polansky, on which basis
further numerical investigations and temperament comparisons are worked out.
%
\subsubsection{Further Well Temperament key characteristics}
From a collection of over one hundred historical temperaments, thirty one were filtered as
WT and included in table \ref{table 8}, based on following criteria:
%
\paragraph{Major thirds}
Almost unanimous historical assessments set the following ratio condition
%
\begin{equation}
1.25<\mathrm{Major\ third} <1.265625\ (=81/64= \mathrm{Pythagorean\ third})
\label{equ 2}
\end{equation}
Compare with: the just third $(5/4=1.25)$ or 12-TET third ($2^{4/12}$=1.2599\dots).
%
\paragraph{Fifths}
Historical assessments of Kirnberger I and II set a lower limit on fifths \citep[][p. 34, lines 17 to 27]{Kelletat1982}\footnote{The Kirnberger II D-A fifth, holding this ratio, was debated by Kirnberger (1721-1783) and
Forkel (1749-1818), and by Sechter (1788-1876), Bruckner (1824-1896).}: the discussed fifth has a ratio of 1.491.
%
\begin{equation}
1.491<fifth <1.509\quad \mathrm{values\ retained\ in\ this\ paper}
\label{equ 3}
\end{equation}
The associated upper limit is derived from the lower limit: it has the same beat.
Some WT have fifth deviations up to one third of a comma. This leads to narrower limits,
not retained here: as many as possible complying WT temperaments are desired.
%
\subsection{Diatonic intervals: \dots ``based on the extended just intonation''; \dots\
\dots ``keep the diatonic intervals as pure as possible'' \dots}
\begin{figure} [h]
\begin{center}
\includegraphics{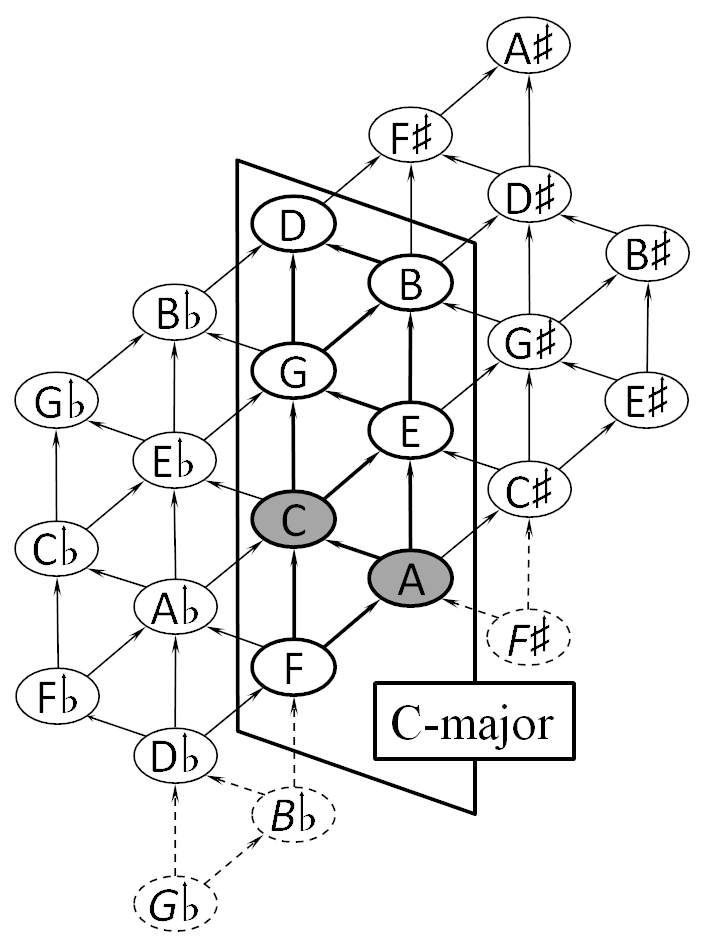}
\caption{C-major diatonic structure}
\label{fig. 2}
\end{center}
\end{figure}

In addition to obligate perfect primes and octaves,
the most important musical intervals are fifths with
perfect ratio of 3/2, major and minor thirds with
just ratio of 5/4 and 6/5 respectively, and their
inversions. These intervals constitute the backbone
of the extended just intonation, building the diatonic C-major
scale, also interconnecting all
altered notes; see fig. \ref{fig. 2}.
%
This structure is not deformable: \textbf{\emph{all triangle sides}} have \textbf{\emph{the fixed
invariable value}} of a \textbf{\emph{perfect}} or \textbf{\emph{just}} interval.

This diatonic C-major scale contains undesired impurity\footnote{Impure interval: an interval deviating from the perfect or just ratio} on the D-F third and the
D-A fifth\footnote{\textbf{\emph{All}} music instruments of the violin family have adjacent D and A strings, and it is common practice to have adjacent strings tuned to perfect fifths; if adjacent strings are major thirds, these are tuned just. See, for example \url{https://en.wikipedia.org/wiki/Standard_tuning}.} (see also footnote 5). There are some more impure intervals on those containing
altered notes. A combination of enharmonics on a single key is wanted for keyboard
instruments. Optimisation of intervals is desirable, \dots and possible.\\
%
%
\subsection{Interval impurity measurement:\ \dots ``keep the diatonic intervals as pure as possible'' \dots}
\begin{figure}
\begin{center}
\includegraphics{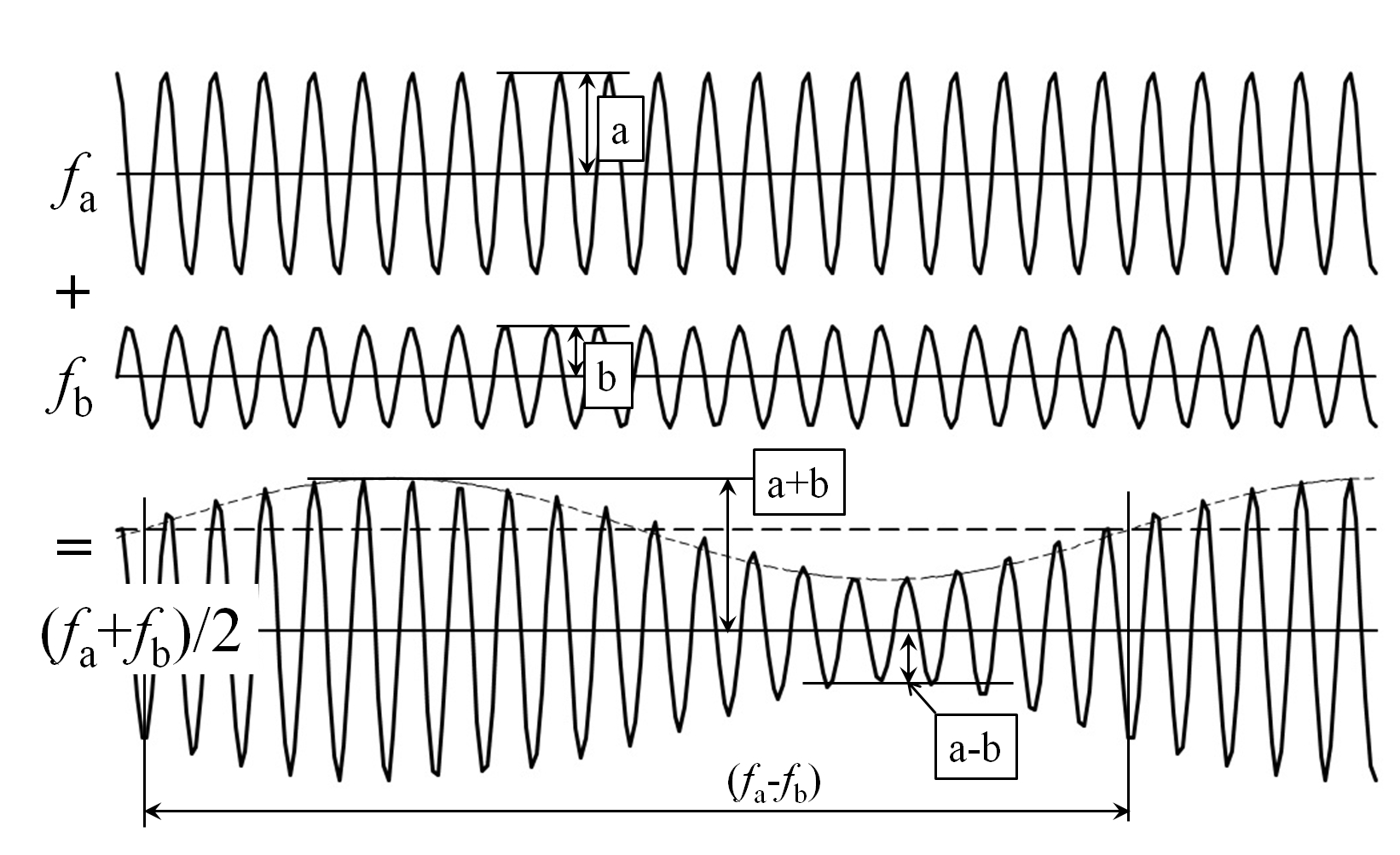}
\caption{Beating of a musical interval}
\label{fig. 3} 
\end{center}
\end{figure}

A mathematical interval impurity measurement is required.\\
Music consists of periodic sounds. J. Fourier (1768-1830) developed mathematical
evidence that any periodic function $F(t)$ consists of a sum of sine waves, -the harmonics-, whereby the sine wave
frequencies are integer multiples of a basic frequency.\\
%
\begin{center}
$F(t)=\sum\limits_{n=0}^{\infty}[a_n\sin{(2\pi nft)}+b_n\sin{(2\pi nft)]}\quad \mathrm{with}\ n=integer\ \mathrm{and\ with:}$
\end{center}
\begin{equation}
a_n=2f \int_{-1/2f}^{1/2f}F(t) \sin(2 \pi nft)dt\quad \mathrm{and}\quad b_n=2f \int_{-1/2f}^{1/2f}F(t) \cos(2 \pi nft)dt
\label{equ 4}
\end{equation}
A small musical interval impurity leads to a beating sound, because of the summation of
mutual note harmonics, of almost
equal frequency. For example: the
beating of an imperfect fifth -ratio\ $\approx 3/2$-, mainly results from the
sum of the second harmonic of the
upper note with the third harmonic
of the lower note, -but also from any
higher harmonic 2n with any mutual higher harmonic 3n-. Fig. \ref{fig. 3} displays the effect.
%
The sum of two sine waves is worked out in formula \ref{equ 5}.
%
\begin{equation}
a\sin(2 \pi f_a t)+b\sin(2 \pi f_b t)=\sqrt{a^2+b^2+2ab\cos2 \pi (f_a-f_b )t} \sin\left (2 \pi \frac{f_a+f_b}{2}t \right)
\label{equ 5}
\end{equation}
Hence, this sum corresponds to a single sine wave:
\begin{itemize}
\item of median frequency $(f_a+f_b)/2$
\item with amplitude varying from $(a-b)$ to $(a+b)$, at modulation frequency $(f_a-f_b)$
\end{itemize}
The lowest beat pitch of a fifth can therefore be set as.
\begin{center}
$Beat_{fifth}=2f_{upper\ note}-3f_{lower\ note}=p_1f_2-p_2f_1$
\end{center}
This impurity measurement has been applied already, among others also by A. Kellner (1977). 
Alternate impurity measurements that can be thought of, are, for example\footnote{$p_1$ and $p_2$ are integers, for example: 
fifths: 3 and 2; major thirds: 5 and 4, minor thirds: 6 and 5.

The factor $p_1$ must be doubled if the upper note corresponding with $f_2$ should lie in the next upper octave.}
%
\begin{eqnarray}
\label{equ 6}
\mathrm{\textbf{\emph{Beat}}}  &=&(p_1 f_2 - p_2f_1)\ \mathrm{cps}\qquad \mathrm{Beat\ Pitch\ (cycles\ per\ second)}\\
\label{equ 7}
\mathrm{\textbf{\emph{PBP}}}  &=&100(p_1 f_2 / f_1-p_2)\ \mathrm{\%}\qquad \mathrm{Percentage\ Beat\ Pitch}\\
\label{equ 8}
\mathrm{\textbf{\emph{PPD}}}  &=&100(p_1 f_2 /p_2f_1-1)\ \mathrm{\%}\qquad \mathrm{Pitch\ Purity\ Deviation}\\
\label{equ 9}
\mathrm{\textbf{\emph{Cents}}}  &=&1200\log(p_1 f_2 /p_2f_1)/\log2\ \mathrm{cents}\qquad \mathrm{Pitch\ Purity\ Deviation\ in\ cents}
\end{eqnarray}
%
\begin{equation}
\mathrm{For\ low\ interval\ impurities\ we\ can\ set:}\qquad p_1 f_2/{p_2 f_1} \approx 1\qquad \mathrm{hence}
\label{equ 10}
\end{equation}
%
\begin{equation}
\log{\left (\frac{p_1 f_2}{p_2 f_1}\right )} \approx \frac{p_1 f_2}{p_2 f_1}-1= \frac{p_1 f_2-p_2 f_1}{p_2f_1} =  \frac{p_1 f_2-p_2 f_1}{C_1^t} =\frac{p_1 f_2/f_1-p_2 }{C_2^t}
\label{equ 11}
\end{equation}
%
\begin{equation}
\mathrm{or}\footnote{Formula \ref{equ 12}: the ``$\sim$'' character signifies: ``almost exactly proportional with''}\qquad Cents  \sim \mathrm{PPD} \sim  \frac{p_1 f_2-p_2 f_1}{p_2f_1}  \sim Beat/C_1^t  \sim PBP/C_2^t
\label{equ 12}
\end{equation}
Impurity measurement differences are of second order only, hence, choice of
measurement can be free: see for example in last paragraph, Appendix B.2.2, table \ref{table B2}, comparing Beat and Cent results. The preferred
measurement for model elaborations in this paper is the Beat. It is an audible and direct
acoustical sound, monitored for aural tuning. Cents will be used for model and temperament comparisons. Comparisons will therefore  
not depend on the chosen diapason, as a result of measurements based on ratios.
%
%
\subsection{Considerations about the application of BEAT Impurity Measurements}

\subsubsection{Dependence on the pitch of the lower note}
The allowed interval beat depends on the lower note pitch.\\
As for aural tuning, it can be said: the higher the lower note of an interval, the
higher the allowed beat pitch \citep[][see: ``C. Discussion'']{Plomp1965}. If more intervals of equal type appear in a formula, the beat pitches
should therefore be adapted by a factor inversely proportional to the pitch of the
lower note, -for example the factor $2^{‒n/12}$-, according table \ref{table 1}.
%
\begin{table} [h]
\begin{tabular}{|l|c|c|c|c|c|c|}
\hline
Note & C & C$\sharp$ & D & E$\flat$ & E & F \\
\hline
Weight value & $2^{-0/12}$ & $2^{-1/12}$ & $2^{-2/12}$ & $2^{-3/12}$ & $2^{-4/12}$ & $2^{-5/12}$ \\
\hline
Weight symbol & $P_C$ & $P_{C\sharp}$ & $P_D$ & $P_{E\flat}$ & $P_E$ & $P_F$ \\
\hline
\hline
Note &  F$\sharp$ & G & G$\sharp$ & A & B$\flat$  & B \\
\hline
Weight value & $2^{-6/12}$ & $2^{-7/12}$ & $2^{-8/12}$ & $2^{-9/12}$ & $2^{-10/12}$ & $2^{-11/12}$ \\
\hline
Weight symbol & $P_{F\sharp}$ & $P_G$ & $P_{G\sharp}$ & $P_A$ & $P_{B\flat}$ & $P_B$ \\
\hline
\end{tabular}
\caption{Note weights applicable with Beat impurity measurements}
\label{table 1}
\end{table}
%

\subsubsection{Obtained precisions of Note Pitches}The proposed note weights, table \ref{table 1}, lead to small errors: these weights deviate from the required exact value, except for the 12TET.\\
This error can be reduced by repeating the calculation, with weights derived from a preceding calculation. Two to three iterations will lead to a pitch precision better than the third decimal digit.\\
The iterations were applied whenever possible. Pitch corrections turned out to be very low, as expected, based on very small differences between proposed weights, table 1, and exact weights.
\subsubsection{Beat - PBP equivalence}The proposed iterations, § 1.4.2, lead to equal pitch values as with PBP measurements, formula \ref{equ 7}. It is an easy way to obtain analytic PBP solutions: the analytic method developed further in § 2.2.1, leads for PBP indeed, to a set of complex non linear equations.
%
\subsection{Dependence on the type of interval}
It is known from musical practice, that perfect fifths capture more attention than just thirds. This can be expressed mathematically, by assigning differing weights to fifths (weight:$ P_{Fi}$), major thirds $(P_{MT})$, and minor thirds $(P_{mt})$. Because of fifths importance, their weights will always be set at $P_{Fi}=1$,  -this weight factor will therefore not appear in formulas-, and the thirds weights will always have to be restricted to $0 \leq P_{MT}$;$P_{mt} \leq 1$.
%
\subsection{Impact of deviations from perfect or just ratios}
Square impurity values offer some advantages for impurity evaluations:
\begin{itemize}
\item Musical
\begin{itemize}
\item Musical impurities above an average impurity have to a certain extent more arithmetic impact, if the square of impurities is applied, see fig. \ref{fig. 4}. This corresponds to some degree with the musical impact on audience \citep{Hall1973}.
\item The square of impurities avoids a distinction between the sign of deviations, as is the case also for musical audience \citep{Plomp1965, Setshares1993}. Attention: the sign is important for aural tuning.
%
\begin{figure} [h]
\begin{center}
\includegraphics{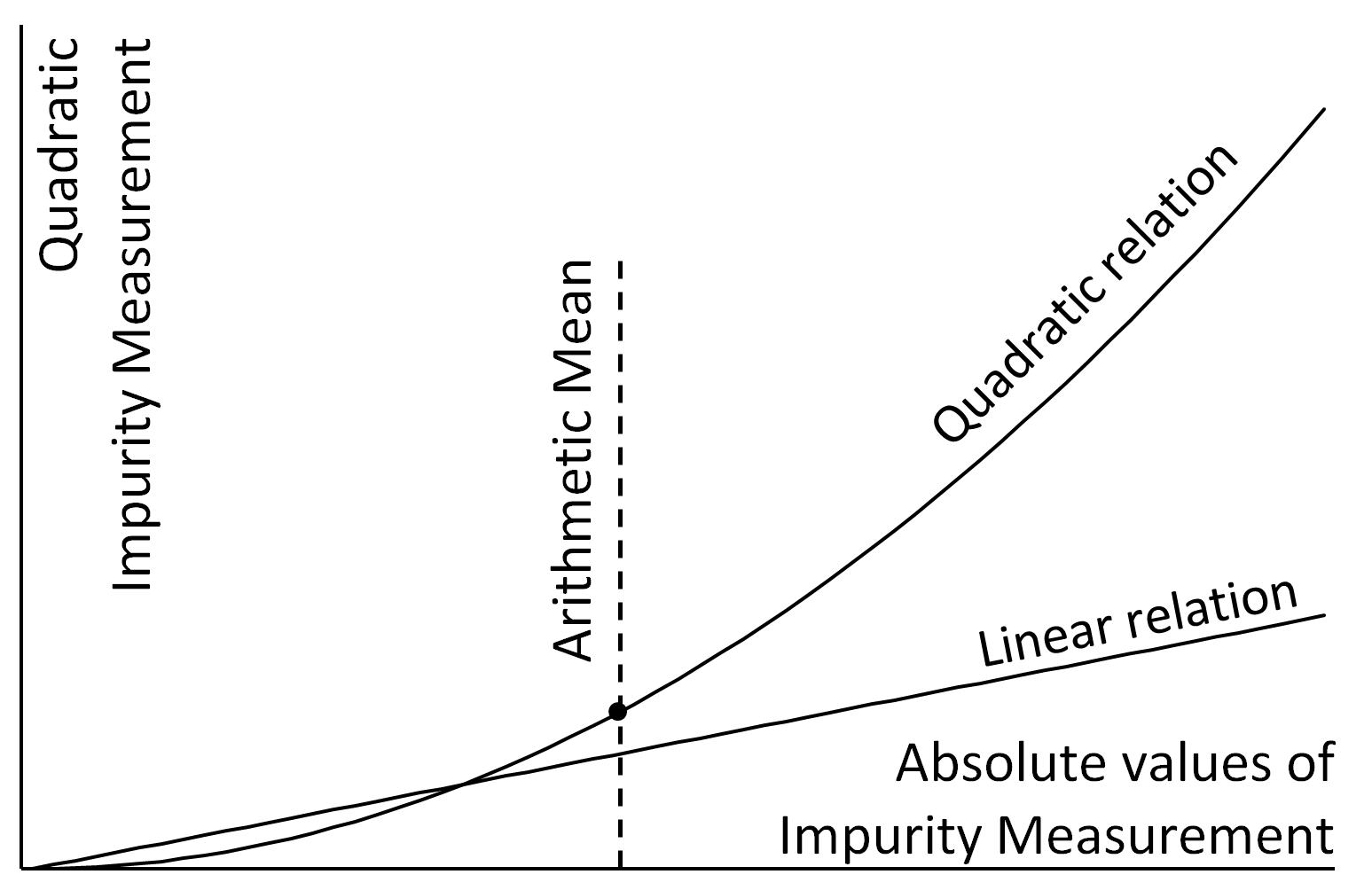}
\caption{Impact of squares}
\label{fig. 4}
\end{center} 
\end{figure}
\end{itemize}
\item Mathematical: for sets of intervals with equal \textbf{\emph{arithmetic}} mean impurity, the -lower
quality- set with larger deviations from the mean, has the largest RMS.
\end{itemize}
Further thoughts on measurements based on interval \textbf{\emph{ratio}} measurement (PBP, PPD, Cents; formulas \ref{equ 7}-\ref{equ 9}): see Appendix B.
%
\section{Elaboration of Optimal Models}
%
\subsection{Impurity Formulas}
Based on concerns of § 1, formulas below can be set ($P_{Note}$: see table \ref{table 1}).
%
\begin{table} [h]
\begin{tabular}{|l|l|}
\hline
Major thirds impurities: & Fifths impurities \\
C: $\Delta MT_C=(4E-5C)P_C$ & C: $\Delta Fi_C=(2G-3C)P_C$ \\
F: $\Delta MT_F=(4A-5F)P_F$ & D: $\Delta Fi_D=(2A-3D)P_D$ \\
G: $\Delta MT_G=(4B-5G)P_G$ & E: $\Delta Fi_E=(2B-3E)P_E$ \\
minor thirds impurities: & F: $\Delta Fi_F=(4C-3F)P_F$ \\
D: $\Delta mt_D=(5F-6D)P_D$ & G: $\Delta Fi_G=(4D-3G)P_G$ \\
E: $\Delta mt_E=(5G-6E)P_E$ & A: $\Delta Fi_A=(4E-3A)P_A$ \\
A: $\Delta mt_A=(10C-6A)P_A$ &  B: A diatonic fifth on B does not exist: \\
B: $\Delta mt_B=(10D-6B)P_B$ & a perfect fifth on B leads indeed to F$\sharp$ \\
\hline
\end{tabular}
\caption{Diatonic C-major impurities}
\end{table}\\
See table 3 for the altered notes, outside the diatonic C-major.
%
\begin{table} [h]
\begin{tabular}{|l|l|}
\hline
C$\sharp$: $\Delta Fi_{C\sharp}=(2G\sharp-3C\sharp)P_{C\sharp}$ & G$\sharp$: $\Delta Fi_{G\sharp}=(4Eb-3G\sharp)P_{G\sharp}$ \\
E$\flat$: $\Delta Fi_{E\flat}=(2B\flat-3E\flat)P_{E\flat}$ & B$\flat$: $ \Delta Fi_{B\flat}=(4F-3B\flat)P_{B\flat}$ \\
F$\sharp$: $\Delta Fi_{F\sharp}=(4C\sharp-3F\sharp G)P_{F\sharp}$ & B: $\Delta Fi_B=(4F\sharp-3B)P_B$ \\
\hline
\end{tabular}
\caption{Impurities of fifths containing an altered note}
\end{table} \\
An optimal diatonic C-major impurity elaboration by minimisation of formula \ref{equ 1}, including \textbf{\emph{preset}} weights $P_{MT}$ and $P_{mt}$, corresponds with the minimisation of
%
\begin{equation}
P_{MT} \sum{\Delta MT_{C,F,G}^2}+P_{mt} \sum{\Delta mt_{D,E,A,B}^2} +\sum{\Delta Fi_{C,D,E,F,G,A}^2}
\label{equ 13}
\end{equation}
and with F and B solved, the minimisation of
%
\begin{equation}
\sum{\Delta Fi_{C\sharp,E\flat,F\sharp,G\sharp,B\flat,B}^2}
\label{equ 14}
\end{equation}
%
%
\subsection{Elaboration of models}
%
\subsubsection{First Model}
A minimum or maximum of formula \ref{equ 13} can be obtained by setting the partial derivatives of the expression to zero, and solving the obtained set of linear equations.\\
Example: the partial derivative by C set to zero, of the expression formula \ref{equ 13}, leads to\\

\begin{figure} [h]
\begin{center}
\includegraphics{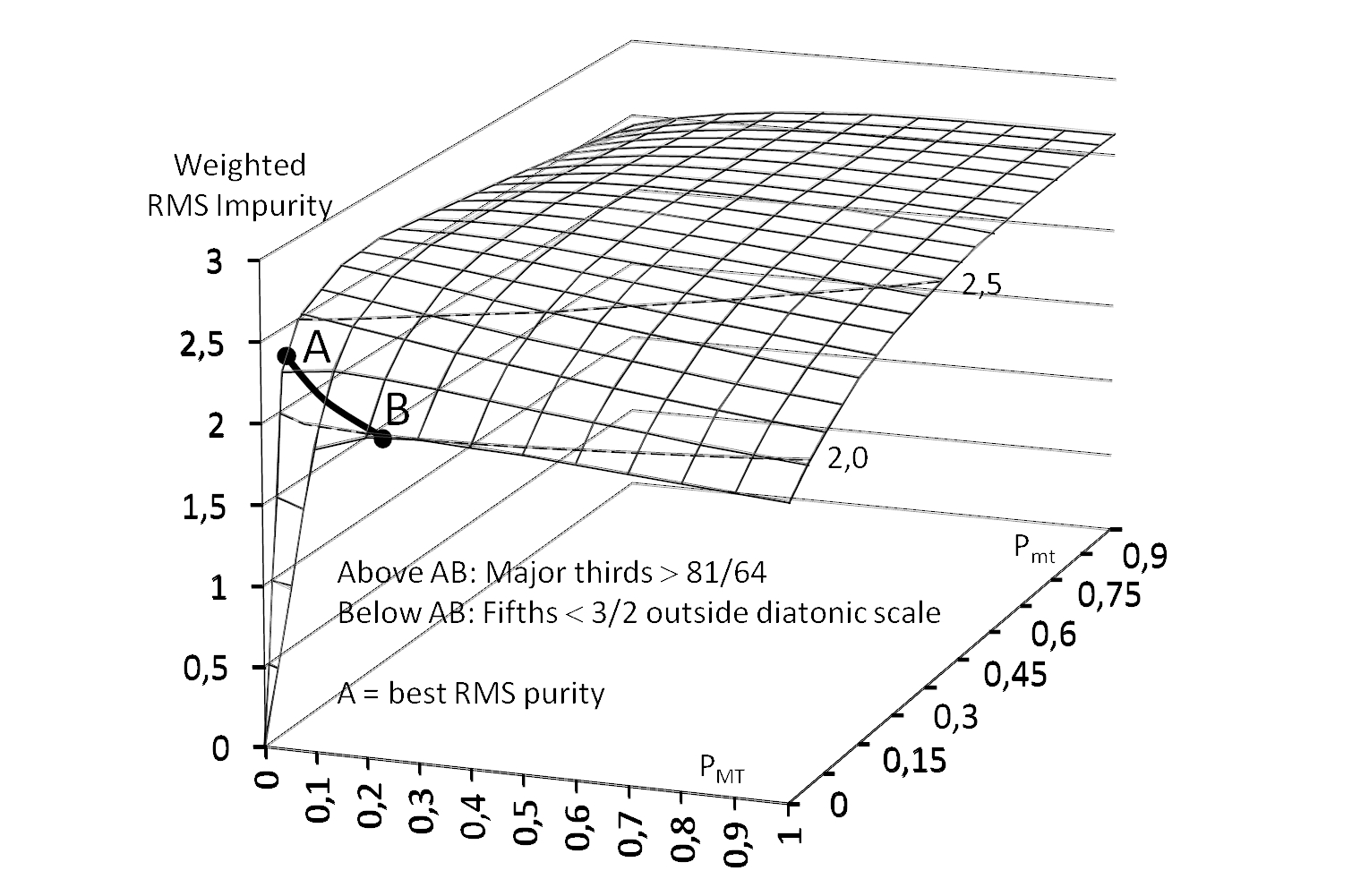}
\caption{Weighted Diatonic Impurity as $F(P_{MT};P_{mt})$}
\label{fig. 5} 
\end{center}
\end{figure}

%
$\frac{\partial}{\partial C} \mathrm{:} \qquad (25P_{MT}P_C^2+100P_{mt}P_A^2+9P_C^2+ 16P_F^2)C$
\begin{equation}
-20P_{MT}P_C^2E-12P_F^2F-6P_C^2G=60P_{mt}P_A^2A
\label{equ 15}
\end{equation}
The full set of equations, including those for altered notes, is available in Appendix A.1, and can be solved by means of a spreadsheet. Separate cells for entry of the $P_{MT}$ 
and $ P_{mt}$  weights are recommended, to enable investigations on their impact. Separate matrices for denominator, and numerators allow for direct calculation and display 
of all results: pitches, impurities, graphs, comparisons, \dots\\
A spreadsheet with the solution of the set of equations is available for downloading.
Fig. \ref{fig. 5} displays the domain of weighted C-major diatonic impurities for $0\leq(P_{MT}; P_{mt})\leq1$.
%
%
\paragraph{Results on the A-B line}
The A-B line separates a WT domain on and below the line, from a non-WT domain holding major thirds in excess of the Pythagorean ratio. $P_{MT}$ and $ P_{mt}$ values in table \ref{table 4} correspond with points on this line. This corresponds with B and F values resulting from six perfect fifths; those on B, F$\sharp$, C$\sharp$, G$\sharp$, E$\flat$, B$\flat$  (thus for $ B/F=2^{10}/3^6$). \\
Point A has the best fifths, and B the best diatonic purity.
%
\begin{table} [h]
\begin{tabular}{|l|c|c|c|c|c|}
\hline
& Point A &&&& Point B\\\hline
 Beat-RMS & 2.364  & 2.266 & 2.172 & 2.077 & 1.988 \\
\hline
Fifths & 1.872 & 1.875 & 1.884 & 1.889& 1.905 \\
\hline
$ P_{mt}$ & 0.05993 & 0.044 & 0.029 & 0.014 &  0.000  \\
\hline
$P_{MT}$ & 0.000 & 0.050 & 0.100 & 0.150 & 0.20364 \\
\hline
\end{tabular}
\caption{Weighted Diatonic Impurity / Unweighted Fifths Impurity}
\label{table 4}
\end{table}
\\
 Models on the A-B line are almost equivalent (see pitches, table \ref{table 5}).
%
\begin{table} [h]
\begin{tabular}{|l|c|c|c|c|c|c|}
\hline
& C & C$\sharp$ & D & E$\flat$ & E & F \\
\hline
\textbf{\emph{Beat-WT-A}} & 262.99 & 277.43 & 294.26 & 312.11 & 329.25 & 351.12 \\
\hline
\textbf{\emph{Beat-WT-B}} & 263.06 & 277.49 & 294.32 & 312.18 & 329.30 & 351.20\\
\hline
\hline
 &  F$\sharp$ & G & G$\sharp$ & A & B$\flat$ & B \\
\hline
\textbf{\emph{Beat-WT-A}} &369.91 & 393.59 & 416.14 & 440.00 & 468.16 & 493.21 \\
\hline
\textbf{\emph{Beat-WT-B}} & 369,99 & 393.76 & 416.24 & 440.00 & 468.27 & 493.32\\
\hline
\end{tabular}
\caption{Beat-WT-A   and   Beat-WT-B}
\label{table 5}
\end{table}
%
\paragraph{Domain below the A-B line $(B/F>2^{10}/3^6)$}
The limit ($P_{MT}=P_{mt}=0$)  has perfect fifths within the diatonic C-major scale, all other fifths being slightly reduced. \textbf{\emph{This ultimate corresponds to a best diatonic scale at altered notes}}, instead of at the diatonic C-major,  \textbf{\emph{which certainly is not the aim}}. The models are WT indeed,  \textbf{\emph{but not optimal}}, since best thirds purities have shifted towards altered notes. This shift can be avoided by imposing the A-B line condition in formula \ref{equ 13}; this is by substituting B by $F\times 2^{10}/3^6$. It opens up new and wider possibilities for exploration of possible diatonic C-major interval optimisation; see further § 2.2.2.
%
\paragraph{Domain above the A-B line $(B/F<2^{10}/3^6)$}
It has been found that WT models within this domain are possible, without inducing major thirds with ratio exceeding the Pythagorean one, by setting specific conditions on some thirds. The corresponding optimal WT model is \textbf{\emph{part of an unexplored WT domain}} 
with extreme diatonic major thirds purity, specific enlarged fifths, and unweighted diatonic purity =3.184. More details are given in Appendix A.3, and explain why this model is not withheld 
as the optimal model in temperament rankings (at tables \ref{table 8} and \ref{table 9}).
%
%
%
\subsubsection{Domain for $B=kF$, with $k=2^{10}/3^6$}
Further diatonic C-major optimisation is possible substituting $B$ by $kF$, and applying the same method as in § 2.2.1. The modified partial derivatives of the sum of the modified squares formula \ref{equ 13}, are given in Appendix A.2 to this paper.

With  F  solved, altered notes can be calculated more easily, based on their perfect fifths ratio, instead of solving the formula \ref{equ 14}.

\begin{figure} [h]
\begin{center}
\includegraphics{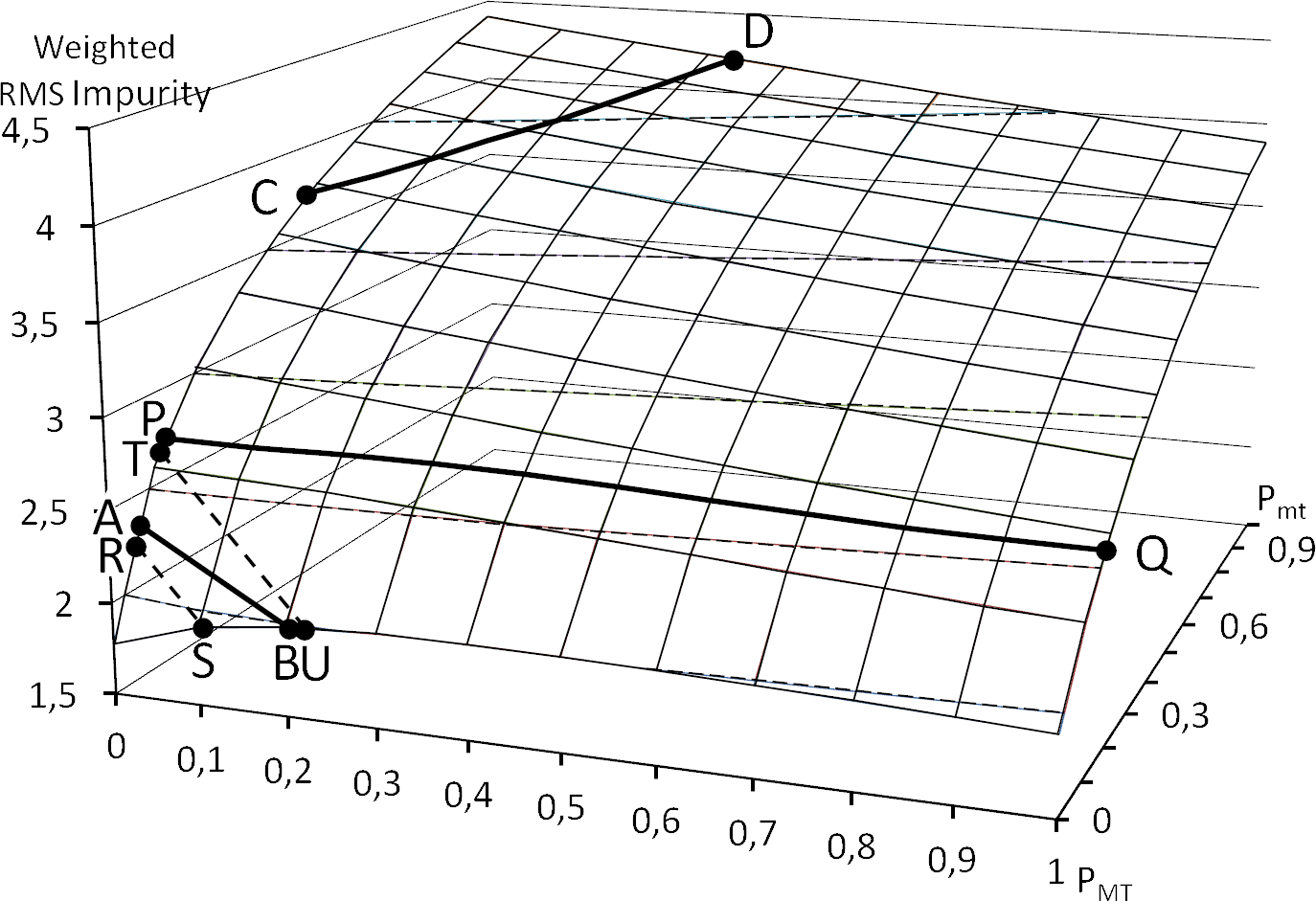}
\caption{Diatonic Impurity as $F(P_{MT}$; $P_{mt}$)}
\label{fig. 6}
\end{center} 
\end{figure}

The domain of solutions is displayed in fig. \ref{fig. 6}. The minimum impurity is obtained at ($P_{MT}=P_{mt}=0$), with a weighted diatonic purity equal to 1.778; this model is identical to the historical Vallotti temperament.
%

A spreadsheet with solution can be downloaded.

\begin{figure} [h]
\begin{center}
\includegraphics{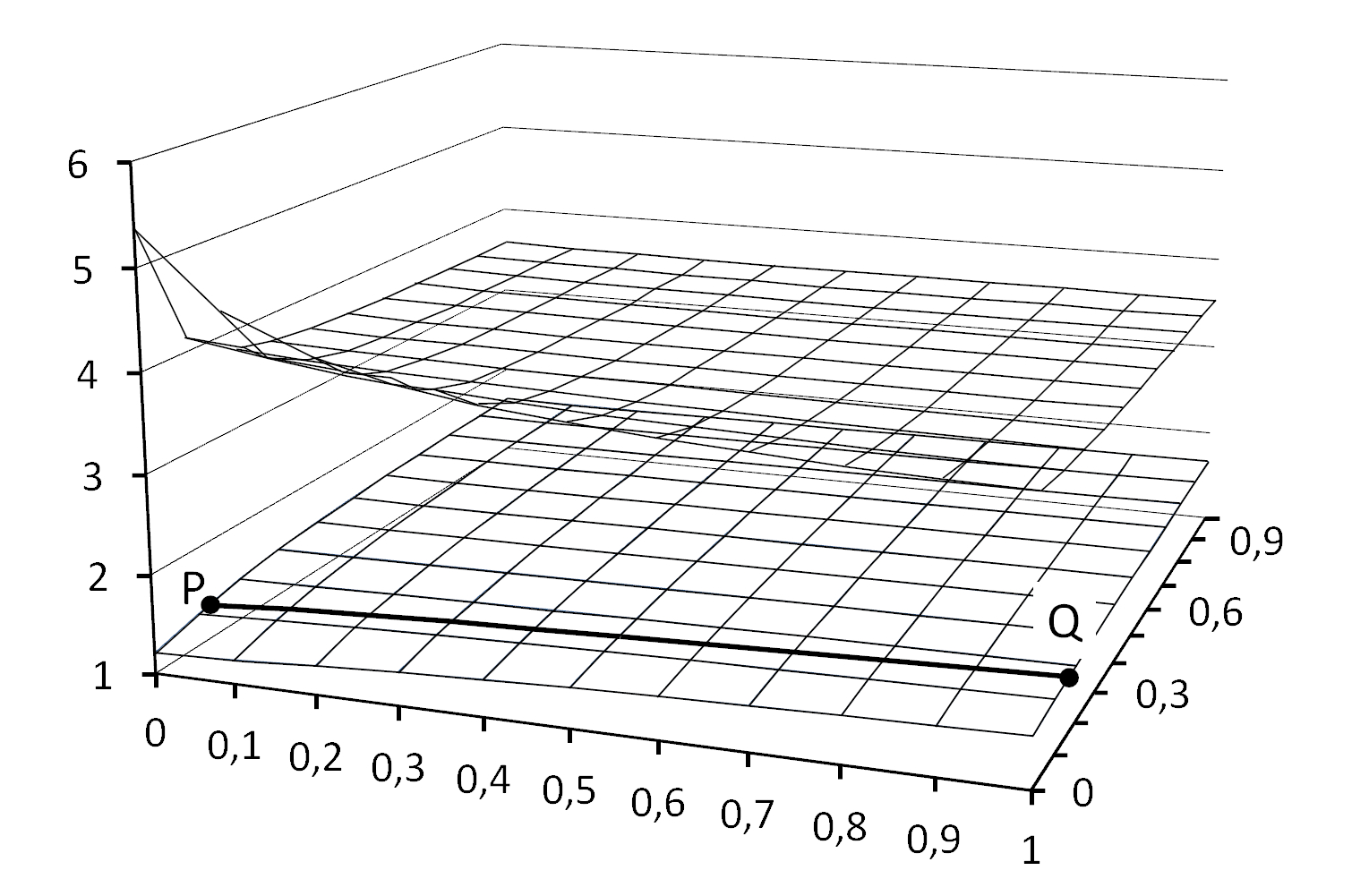}
\caption{Course of unweighted diatonic thirds and fifths}
\label{fig. 7}
\end{center}
\end{figure}

Plenty of combinations of $P_{MT}$ and $P_{mt}$ are possible, but solutions above the C-D line, for ($P_{MT}$; $P_{mt}$) from (0; 0.466) to (0.357; 1), are not WT, because of major thirds exceeding the 81/64 ratio. The almost free choice of $P_{MT}$ and $P_{mt}$ values allows for investigation on which combination might be preferred, taking additional considerations into account, on top of the calculated and desired optimal diatonic C-major impurity. Following musical considerations make sense:
\begin{itemize}
\item Maintain the  \textbf{\emph{C-major diatonic fifths}} as close as possible to the perfect ratio
\item Achieve a good balance between thirds and fifths impurities
\end{itemize}
These two points above, require insight into the course of the impurity contributions of thirds and fifths.\\
Those  \textbf{\emph{group}} contributions can be investigated by means of formulas \ref{equ 16}.
%
\begin{equation}
RMS_{MT;mt}=\sqrt{\frac{\sum{(\Delta MT_{C,F,G}^2+\Delta mt_{D,E,A,B}^2)}}{13}} \mathrm{;}\ RMS_{Fi}=\sqrt{\frac{\sum{\Delta Fi_{C,D,E,F,G,A}^2}}{13}}
\label{equ 16}
\end{equation}
The thirds impurities have a monotone continuous decrease from point $(P_{MT}=P_{mt}=0)$ to $(P_{MT}=P_{mt}=1)$; the fifths have an opposite evolution; see fig \ref{fig. 7}.
%
\\ Because of this opposition, some kind of balance should be possible between those two points, but lack of intersection of planes in fig. \ref{fig. 7} reveals that some weighting of fifths and thirds impurity contributions is necessary.

\begin{figure} [h]
\begin{center}
\includegraphics{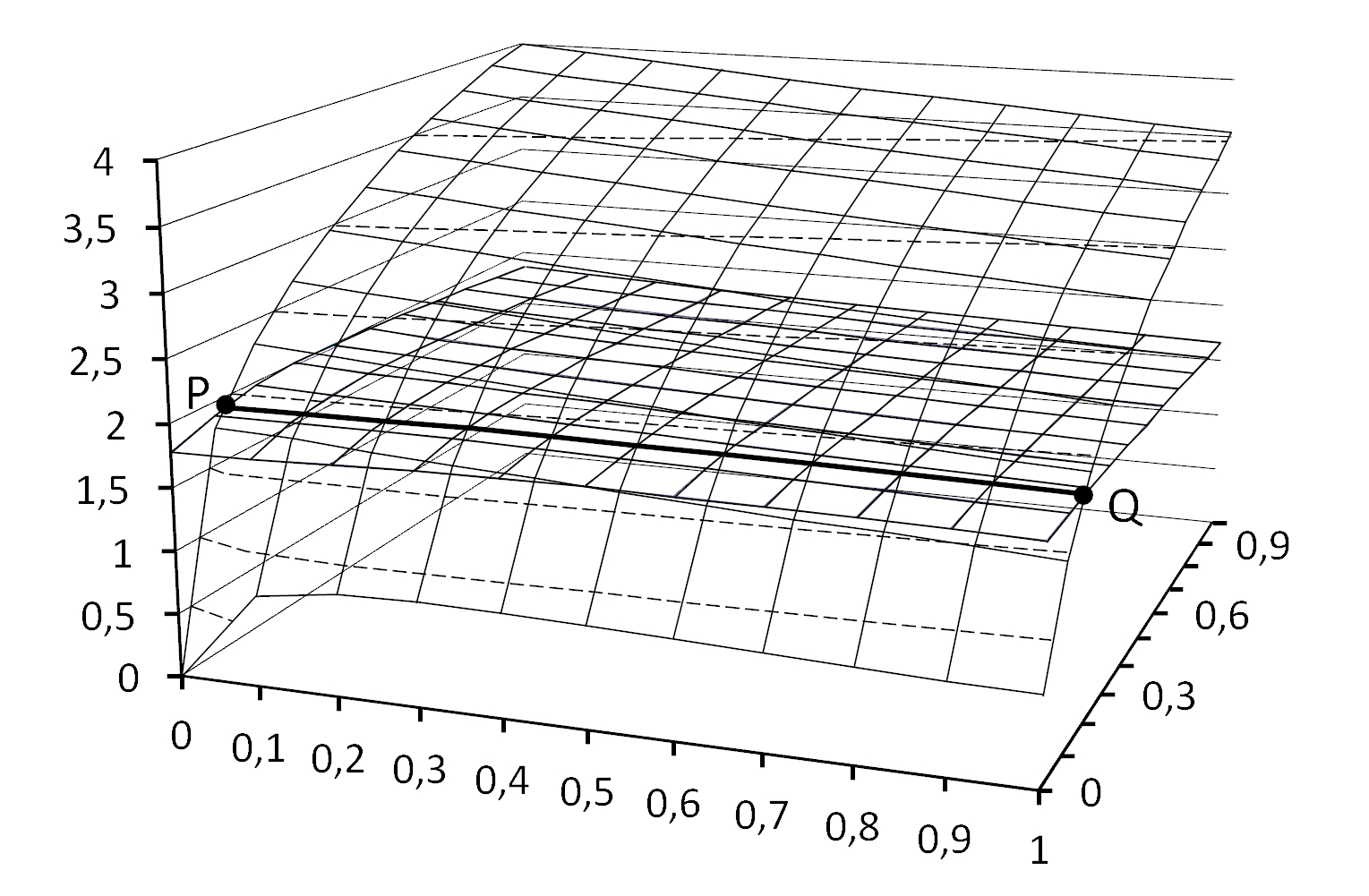}
\caption{Course of weighted diatonic thirds and fifths} 
\label{fig. 8}
\end{center}
\end{figure}

A weighting balance is proposed, by introducing weights in formula \ref{equ 16}; this corresponds with a dissociation of thirds and fifths contributions within formula \ref{equ 1}; see formula \ref{equ 17}.
%
\begin{equation}
\sqrt{\frac{P_{MT} \times \sum{\Delta MT_{C,F,G}^2}+P_{mt} \times \sum{\Delta mt_{D,E,A,B}^2}}{3P_{MT}+4P_{mt}+6}}=\sqrt{\frac{\sum{\Delta Fi_{C,D,E,F,G,A}^2}}{3P_{MT}+4P_{mt}+6}}
\label{equ 17}
\end{equation}
Any other balancing proposal based on rational considerations and objectivity should be acceptable\footnote{For example: setting of weights so that the purity of the best (minor or major) third should not be better than that of the worst fifth was investigated, and did lead to the present proposal: the present balance leads to a model with intermediate $\Delta Fi$: 
(best $|\Delta MT|=0.86)<\ $(worst $|\Delta Fi|=2.86)<\ $(best $|\Delta mt|=4.60)$.}.\\
The course of weighted thirds and fifths purities, based on the proposal of formula \ref{equ 17}, is displayed in fig. \ref{fig. 8}.

%
\noindent The most curved plane displays the thirds impurities. The P-Q line displays the collection of weights leading to equal impurity contribution of thirds and fifths (this line is also displayed in figures \ref{fig. 6} and \ref{fig. 7}). Some numeric figures corresponding with this P-Q line are displayed in table \ref{table 6}.

%
\begin{table} [h]
\begin{tabular}{|l|c|c|c|c|c|c|}
\hline
\textbf{\emph{Fifths Impurity}} & 1.3481 & 1.4221 & 1.4704 & 1.5059 & 1.5334 & 1.5555 \\
\hline
Diatonic Impurity & 4.2928 & 4.1030 &  4.0257 & 3.9859 & 3.9630 & 3.9488 \\
\hline
Weighted Diatonic Impurity & 2.7024 &  2.7148 &  2.6874 & 2.6457 & 2.5982 & 2.5486 \\
\hline
$P_{mt}$ & 0.1177 &  0.1335 & 0.1457 & 0.1557 & 0.1640 & 0.1710 \\
\hline
$P_{MT}$ & 0.0 & 0.2 &  0.4 &  0.6 & 0.8 & 1.0 \\
\hline
\end{tabular}
\caption{Impurities of Fifths and the C-major diatonic scale}
\label{table 6}
\end{table}
\noindent A best fifths purity combined with a thirds to fifths balance, can be observed at point P (for $ P_{MT}=0$ table \ref{table 6}). The best fifths impurity is 13.3\% smaller than the worst one, for a diatonic impurity increase of 8.7\% only (6\% weighted). Because of fifths purity importance, the point P is preferred: \textbf{\emph{fifths purity}} is \textbf{\emph{directly audible}}, indeed; \textbf{\emph{musical evaluation}} of \textbf{\emph{thirds or diatonic purity}} is \textbf{\emph{difficult}} (see further, footnote 12, items (1) and (3)).
Pitches of this model, named Beat-WT, are on display in table 7.
%
\begin{table} [h]
\begin{tabular}{|l|c|c|c|c|c|r|}
\hline
\textbf{\emph{Beat-WT}} & C & C$\sharp$ & D & E$\flat$ & E & F \\
\hline
Piches & 263.23 & 277.55 & 294.39 & 312.25 & 329.25 & 351.28 \\
\hline
\hline
\textbf{\emph{Beat-WT}} &  F$\sharp$ & G & G$\sharp$ & A & B$\flat$ & B \\
\hline
Pitches & 370.07 & 393.95 & 416.33 & 440.00 & 468.37 & 493.43 \\
\hline
\end{tabular}
\caption{impurity = 2.7024; (4.2928 unweighted)	Beat-WT	$P_{MT}=0$; $P_{mt}=0.1177$}
\label{table 7}
\end{table}
%
%
\section{Evaluation and Comparison of Models and Historical WT}
Most data on historical temperaments were obtained from Kelletat \citeyearpar{ Kelletat1981,  Kelletat1982} and De Bie \citeyearpar{De Bie2001}; or were self computed: 
Kirnberger III unequal for example \citep[][see data on fig. 12]{Kelletat1982}. A warning from E. Amiot \citeyearpar[][§ {[2.1.5]}]{Amiot2009} applies here also: ``I hope that through the process of chain-quotation, the exact values of the tunings have been preserved, \dots''.
\subsection{Purity Ranking}
Purity ranking depends on assigned $P_{MT}$ and $P_{mt}$ weights. Diatonic C-major impurities of models and historic temperaments have been controlled, for the full collection ($ 0\leq P_{MT};P_{mt} \leq 1$). It was almost always perceived that the Beat-WT model and Kirnberger III are ranked at the top, with the 12TET on the bottom.
\subsubsection{Unweighted diatonic C-major impurities $(P_{MT}=P_{mt}=1)$}
The Beat-WT model has the best unweighted impurity, except for Vogel: but Vogel has peculiar fifths properties (with $B<2^{10}/3^{6}F$) leading to better thirds; (see § 2.2.1.3, and appendix A3, fig. A1), and is as such, part of an unexplored WT domain.
\subsubsection{Weighted diatonic C-major impurities $(0\leq P_{MT}$;$P_{mt}<1)$.}
The Beat models always have the best weighted impurity, as mathematically expected, and Kirnberger III is always on top,  \textbf{\emph{except if $P_{MT}$ AND $P_{mt}$ have very low values}}\footnote{See fig. 6: Beat-WT is best as soon as ($P_{MT}$; $P_{mt}$) lies above the R-S line [(0; 0.045) to (0.103; 0)].
Kirnberger III is the best historic WT as soon as ($P_{MT}$; $P_{mt}$) lies above the T-U line [(0; 0.117) to (0.213; 0)].
}: around the extreme $(P_{MT}=P_{mt}=0)$, quite a lot of historic temperaments can then have better diatonic C-major purity, as consequence of wider diatonic fifths (with $B>2^{10}/3^{6}F$), but with lack therefore, of some optimality of thirds (see § 2.2.1).\

\noindent Based on the above, the most relevant ($P_{MT};P{mt}$) combinations are on display in table \ref{table 8}: unweighted impurities, and weighted impurities based on quite low weights matching the Beat-WT model (table 7).\\
%
\begin{table} [h]
\begin{tabular}{|l|c|l|r|}
\hline \qquad \textbf{Unweigthed impurities} & & \qquad \textbf{Weigthed impurities} & \\
\hline  \quad \qquad $P_{MT}=P_{mt}=1$ & & \quad  $P_{MT}=0$;  $ P_{mt}=0.1177$  & \\
\hline Vogel (Stade) 1975 & 3.997 & \textbf{\emph{Beat-WT}} & 2.702  \\
\hline \textbf{\emph{Beat-WT}} & 4.291 &  Vogel (Stade) 1975 & 2.706  \\
\hline \textbf{\emph{Beat-WT-B}} & 4.508 & \textbf{\emph{Beat-WT-B}} & 2.717 \\
\hline \textbf{\emph{Beat-WT-A}} & 4.623 & \textbf{\emph{Beat-WT-A}} & 2.731  \\
\hline (NK)\quad  Kirnberger III ungl. & 4.737 & (NK)\quad  Kirnberger III 1779 & 2.847 \\
\hline (NK)\quad  Kirnberger III 1779 & 4.807 & (NK)\quad  Kirnberger III ungl. & 2.865  \\
\hline (K)\quad  Kelletat 1966  & 4.839 & Sievers 1868 & 2.866  \\
\hline Stanhope 1806 & 4.929 & (K)\quad  Kelletat 1966 & 2.921 \\
\hline Sievers 1868 & 4.972 & (N)\quad Vallotti-Tartini 1750 & 2.941  \\
\hline (N)\quad Vallotti-Tartini 1750 & 5.499 & (K)\quad Kellner 1976 & 2.980  \\
\hline (K)\quad Kellner 1976 & 5.684 & (N)\quad Young 1800 & 3.029  \\
\hline (K)\quad Billeter 1979 & 5.874 & Mercadier 1788 & 3.048  \\
\hline (N)\quad Werckmeister III 1681 & 6.000 & Barca (Devie) 1786 & 3.061  \\
\hline Barca (Devie) 1786 & 6.107 & Young / Van Biezen 1975 & 3.087  \\
\hline (N)\quad Young 1800 & 6.129 & Barnes 1979 & 3.087 \\
\hline (N)\quad  Neidhardt-4 1732 & 6.130 & Lehman 2005 & 3.087  \\
\hline Young / Van Biezen 1975 & 6.205 & Stanhope 1806 & 3.090  \\
\hline Barnes 1979 & 6.205 & Neidhardt-1 1724 & 3.029  \\
\hline Lehman 2005 & 6.205 & (N)\quad  Neidhardt-4 1732 & 3.135  \\
\hline Mercadier 1788 & 6.215 & Lambert 1774 & 3.159  \\
\hline Neidhardt-1 1724 & 6.227 & (K)\quad Billeter 1979 & 3.162  \\
\hline Lambert 1774 & 6.551 & (N)\quad Werckmeister III 1681 & 3.178  \\
\hline Barca (Asselin) 1786 & 6.753 &  Barca (Asselin) 1786 & 3.209 \\
\hline Bendeler II 1690 & 7.159 & Neidhardt-2 1724 & 3.302 \\
\hline Neidhardt-2 1724 & 7.163 & Asselin 1985 & 3.356  \\
\hline Asselin 1985 & 7.363 & Sorge 1744 & 3.394  \\
\hline Sorge 1758 & 7.434 & Neidhardt-3 1724 & 3.394  \\
\hline Sorge 1744 & 7.477 & Sorge 1758 & 3.431  \\
\hline Neidhardt-3 1724 & 7.477 & Bendeler II 1690 & 3.586 \\
\hline Werckmeister V 1681 & 7.548 & Werckmeister V 1681 & 3.655 \\
\hline Bendeler III 1690 & 7.574 & Bendeler III 1690 & 3.667 \\
\hline Gothel 1855 & 8.013 & Fritz 1756 & 3.841 \\
\hline Fritz 1756 & 9.091 & Gothel 1855& 3.863 \\
\hline 12TET/Galileo 1581 & 9.304 & 12TET/Galileo 1581 & 3.902 \\
\hline
\end{tabular}
\caption{Diatonic WT Impurities: C-major impurities}
\label{table 8}
\end{table}
\noindent Kirnberger III, Kirnberger III ungleich, Vallotti, Werckmeister III, Neidhardt, Young, in frames marked by (N) in table \ref{table 8}, are installed on more than 65\% of the organs of the Netherlands, according statistics of the Huygens-Fokker Foundation database, the Netherlands; \url{http://www.huygens-fokker.org/docs/orgeltemp.html}. Most of the remaining organs are tuned in meantone ($>20\%$, non-WT), two are tuned close to 12TET.

Kelletat  claims equivalence for temperaments in frames marked by (K) in table \ref{table 8}. \textbf{\emph{It should not be possible to notice an audible difference}}  \citeyearpar[][p. 142]{Kelletat1982}\footnote{\begin{enumerate}\item Jede von diesen Temperaturen (= temperaments marked by (K) in table \ref{table 8}) ist praktisch gans gleich richtig oder gleich falsch. \item {\dots} \item Wer solche Überlegungen weiter nicht nötig hat, kann sich die Temperatur aussuchen, die am einfachsten zu legen ist. Das ist mit großem Abstand die nach Kirnberger III mit ungleicher Teilung der Kommas.\textbf{\emph{ Und er kann, nach vielfältiger Erfahrung, ganz sicher sein, daß beim Anhören von Musik niemand diese Temperatur von einer anderen der hier dargestellten unterscheiden kann.}} \end{enumerate}}:
\begin{quote}
(1) The sound of any one of those temperaments (= temperaments in frames marked by (K) in table 8) is practically equally good or equally false\\
(2) \dots\\
(3) Those who do not need such considerations can choose the temperament that is easiest to tune. This is by far the one after Kirnberger III ungleich with unequal division of the commas. \textbf{\emph{And they can, after many experiences, be quite sure that when listening to music, no one can distinguish this temperament from any of the others presented here.}}
\end{quote}The above cited temperaments, marked by (N) or (K) in table \ref{table 8},  have a favourable ranking.

\subsection{Comparisons with the Beat-WT model}

Another ranking is possible, \textbf{\emph{comparing}} historical temperaments with the obtained Beat-WT model. This comparison makes sense: a low C-major diatonic impurity does not necessarily imply best possible fifths in good balance with thirds, as was indicated during the development of the Beat-WT model (§ 2.2.2). Temperaments can be \textbf{\emph{compared}}, by calculating the RMS-$\Delta$-Cents of \textbf{\emph{mutual impurity differences}} of \textbf{\emph{all thirds and fifths}}, applying formula \ref{equ 18} (with Cent impurities according formula \ref{equ 9}).\\

%
\noindent $RMS$-$\Delta$-$Cents_{C; C\sharp; D; E\flat; E; F; F\sharp; G; G\sharp; A; B\flat; B} $
\begin{equation}
=\sqrt{\frac{\sum{[(\Delta MT_1 - \Delta MT_2)^2+(\Delta mt_1 - \Delta mt_2)^2+(\Delta Fi_1 - \Delta Fi_2)^2]}}{36}}
\label{equ 18}
\end{equation}
This formula is useful too, for very precise and accurate \textbf{\emph{recognition of temperaments}}.
\textbf{\emph{Recognition}} can be carried out by comparing the unknown temperament with a large collection of historical temperaments, also those that are not WT. The unknown temperament is almost certainly the historical temperament corresponding with the lowest RMS-$\Delta$-Cents difference.

Results of comparisons are displayed in table \ref{table 9}. The frames are marked in the same way as table \ref{table 8}, and a favourable ranking of most marked temperaments can again be observed. Neidhardt, Werckmeister, and Billeter have shifted slightly downwards.

%
\begin{table} [h]
\begin{tabular}{|r|c|l|c|c|c|l|}
\hline 0 & 0.00 & \textbf{\emph{Beat-WT}} & & 16 & 2.97 & (K)\quad  Billeter 1979  \\
\hline 1 & 1.17 & (NK)\quad Kirnberger III 1779& & 17 & 3.08 & Lehman 2005  \\
\hline2 & 1.45 & (NK)\quad Kirnberger III ungl.  & & 18 & 3.10 & Lambert 1774 \\
\hline 3 & 1.87 & Sievers 1868 & & 19 & 3.28 & Young (Van Biezen) 1975  \\
\hline 4 & 2.00 & (N)\quad Vallotti - Tartini 1750 & & 20 & 3.40 & Barca (Asselin) 1786 \\
\hline 5 & 2.18 & (K)\quad Kelletat 1966 & & 21 & 3.56 & Neidhardt-2 1724 \\
\hline 6 & 2.24 & Vogel (Stade) 1975 & & 22 & 3.67 & Asselin 1985  \\
\hline 7 & 2.30 & (N)\quad Young 1800 & & 23 & 3.81 & Sorge 1744  \\
\hline 8 & 2.35 & (K)\quad Kellner 1976 & & 24 & 3.92 & Neidhardt-3 1724  \\
\hline 9 & 2.49 & Stanhope 1806 & & 25 & 3.92 & Sorge 1758  \\
\hline 10 & 2.56 & Barca (Devie) 1786 & & 26 & 4.52 & Bendeler-II 1690  \\
\hline 11 & 2.56 & Barnes 1979 & & 27 & 4.66 & Bendeler III 1690  \\
\hline 12 & 2.57 & Mercadier 1788 & & 28 & 5.08 & Gothel 1855  \\
\hline 13 & 2.58 & (N)\quad  Neidhardt-4 1732 & & 29 & 5.26 & Werckmeister V 1681  \\
\hline 14 & 2.77 & (N)\quad Werckmeister III 1681 & & 30 & 5.64 & Fritz 1756  \\
\hline 15 & 2.88 & Neidhardt-1 1724 & & 31 & 5.78 & 12TET/Galileo 1581  \\
\hline
\end{tabular}
\caption{$RMS$-$\Delta$-$Cents$ comparison of historical WT temperaments with Beat-WT model}
\label{table 9}
\end{table}
Similar rankings can be obtained by sorting with a customised maximum MSS \citep{Amiot2009}\footnote{Define the MSS of the Discrete Fourier Cross Correlation between two temperaments, instead of the MSS of the Discrete Fourier Transform} or a customised $\delta$ \citep[][§ 5]{Liern2015}\footnote{Define $\delta$ by adapting the diapason of temperament $S_2$ for either a minimised maximum $d(s_i, t_i)$, or a minimum $RMS[d(s_i, t_i)]$}; all comparisons have Kirnberger III at the top and 12TET at the bottom. This high similarity is striking: it strongly supports a hypothetical recommendation to prefer the Kirnberger III temperament for general musical practice, except for some specific Baroque music (see further § 4).
 \subsection{Kirnberger III}
Kirnberger attaches great importance to purity in general \citeyearpar[][part I, Preface]{Kirnberger}\footnote{ (Vorrede)\dots Ueberall habe ich die höchste Reinigkeit zum Augenmerk gehabt, weil ich gefunden habe, daß die größten Genie in die Composition, sie sorgfültig gezucht haben. \dots\\
\dots Aber ich weiß es aus langer Erfahrung, wie nützlich es ist, die angehenden Componisten an die strengste Reinigkeit zu gewöhnen. \dots}:
\begin{quote}
\dots  Everywhere I strived for the utmost purity, because I have found that the greatest geniuses in composition have cared for it carefully. \dots\\
\dots But I know from long experience how valuable it is to accustom the aspiring composers to the strictest purity. \dots
\end{quote}
He attaches great importance to thirds purities \citeyearpar[][part II, p. 71]{Kirnberger}\footnote{Man kann als eine Grundregel zu Beurtheilung der Tonleiter annehmen, daß die Durtöne, deren Terzen ganz rein sind, die der Durtonart zukommende Eigenschaft am vorzüglichsten bezißen, und daß in die Durtöne, die sich am meisten von dieser Reinigkeit entfernen, auch am meisten Rauhigkeit und zulezt etwas von Wildheit komme.}:
\begin{quote}
As a basic rule of judgment of tonalities, it can be assumed that major tonalities, with quite pure thirds, are the most characteristic ones, and that the major tonalities, which are most remote from this pureness, are also the roughest ones that comes from some wildness.
\end{quote}
He proposes a just major third on C, on top of the desired diatonic C-major purity \citep[][p. 47, letter of Kirnberger to Forkel, 1779]{Kelletat1981}\footnote{\dots wenn man will, lasse man C-E ganz rein und stimme diese vier Quinten C-G, G-D, D-A, A-E, jede Quinte abwartsschwebend,\dots}: "\dots if wanted, C-E can be tuned just, and these four fifths, C-G, G-D, D-A, A-E, each tuned slightly below perfect,\dots".

Models based on the algorithms of this paper, with a preset of a just major third on C, are worked out in appendix B3. The best fit with Kirnberger III is obtained with optimisation of fifths only ($P_{MT}=P_{mt}=0$), it has a $RMS$-$\Delta$-$Cent=0.945$, and a diatonic C-major impurity of 4.41.\\

Farup too developed an optimised mathematical model with a just major third on C \citeyearpar[][the model on top of fig. 5]{Farup2013}, it fits Kirnberger III 
with a $RMS$-$\Delta$-$Cent=0.62$, and has a diatonic C-major impurity of 4.68.
 \subsection{Further evaluations}
\begin{figure} [h]
\begin{center}
\includegraphics{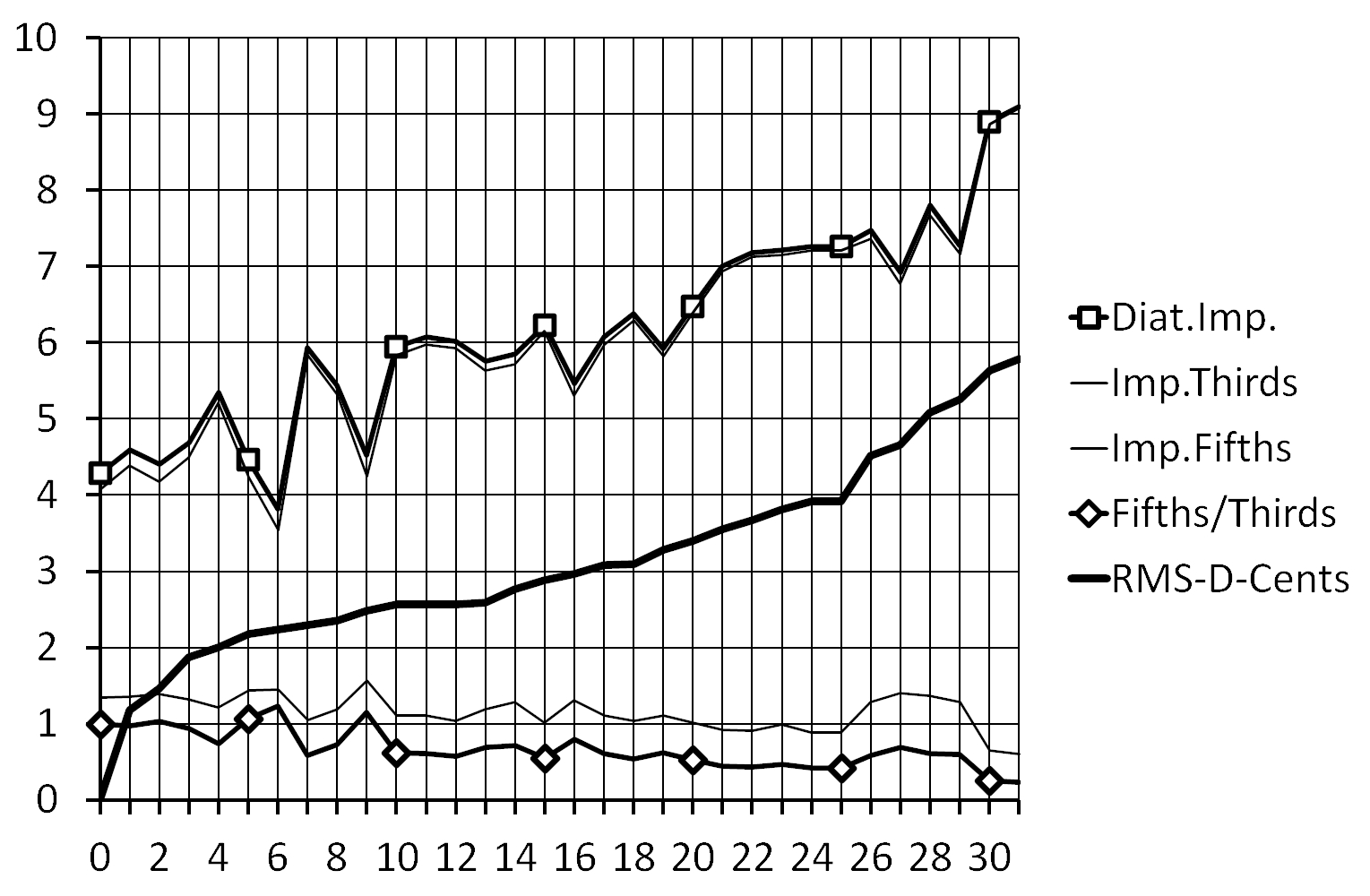}
\caption{Historical WT characteristics ranked according table \ref{table 9}} 
\label{fig. 9}
\end{center}
\end{figure}

The courses of RMS-$\Delta$-cents, of  the \textbf{\emph{unweighted}} diatonic C-major, its fifths and thirds impurities, and the \textbf{\emph{weighted}} Fifths/Thirds impurity ratios, against the ranking in table \ref{table 9} are displayed in fig. \ref{fig. 9}.

%
\begin{figure} [h]
\begin{center}
\includegraphics{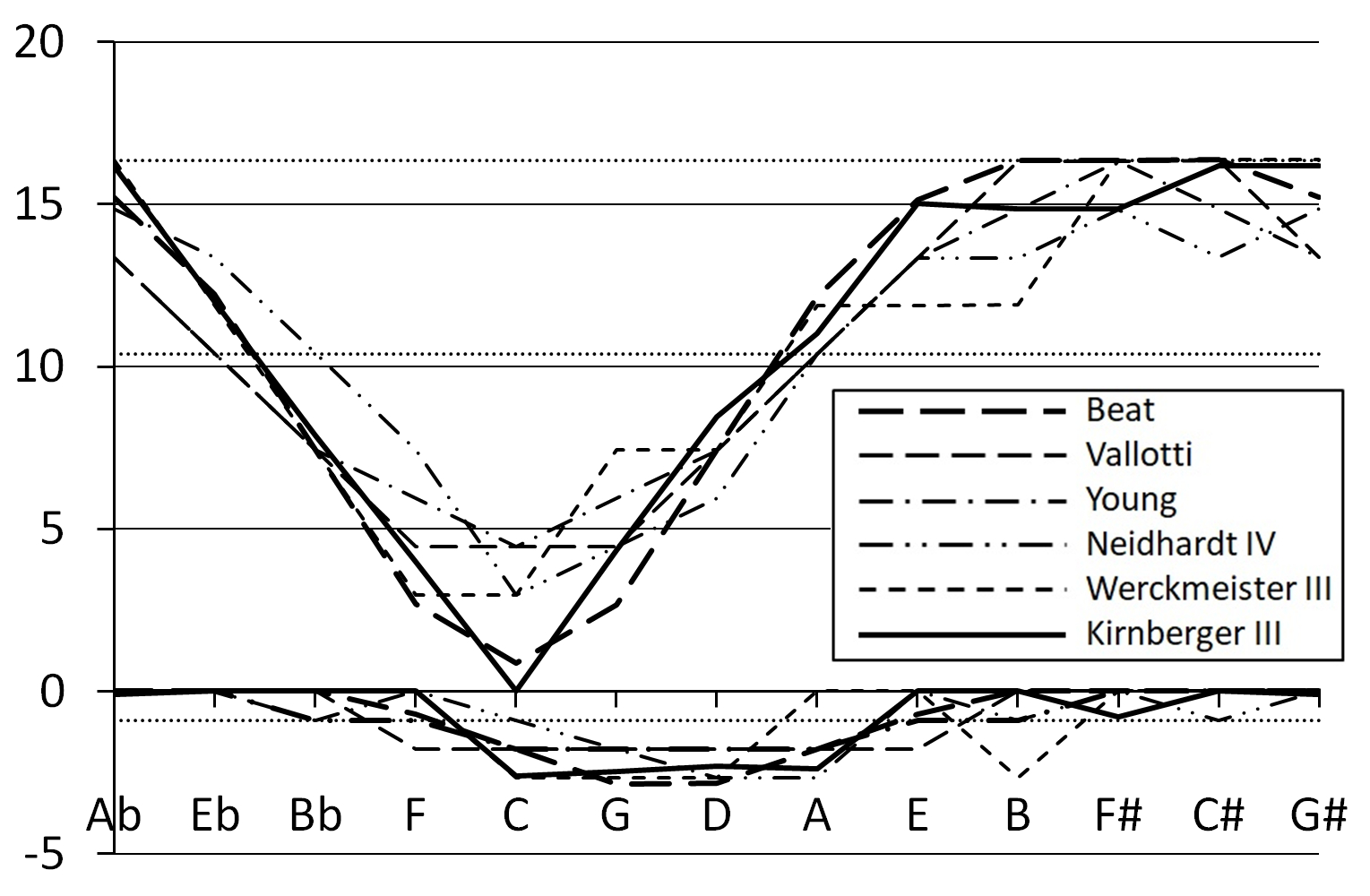}
\caption{Beat impurity of major thirds (up) and fifths (down)} 
\label{fig. 10}
%
\includegraphics{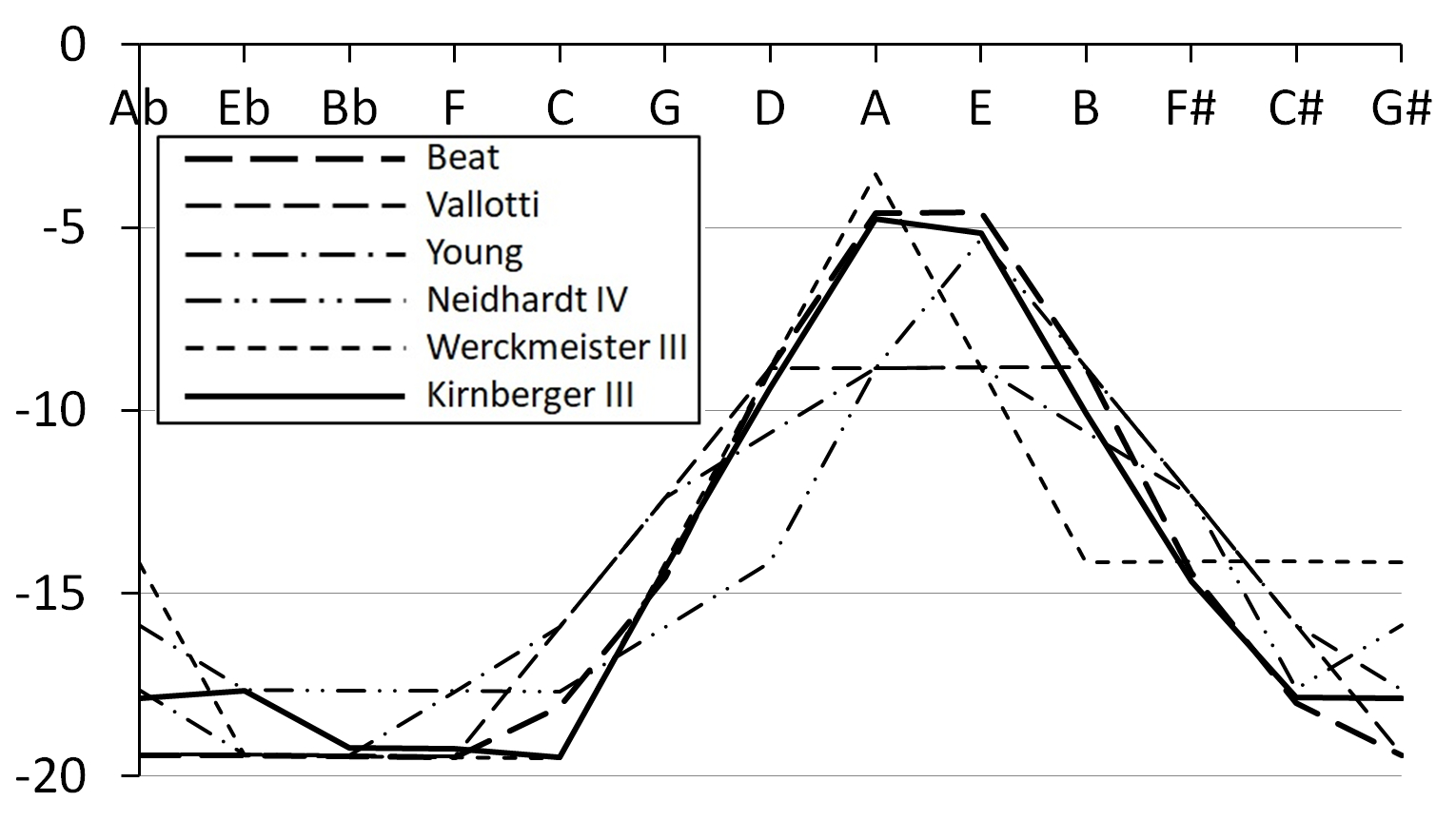}
\caption{Beat impurity of minor thirds}
\label{fig. 11} 
\end{center}
\end{figure}

\noindent Notice the low diatonic C-major impurity at position 6 (Vogel).\\
 The C-major diatonic impurity is only slightly higher than the thirds impurity. It should be noticed here that many publications duly attach much attention indeed to \textbf{\emph{major}} thirds quality. But it is striking on the other hand, that the Beat-WT model is determined by the weights of \textbf{\emph{minor thirds only}} ($P_{MT}=0$; $P_{mt}=0.1177$; tables \ref{table 6}, \ref{table 7}). Best mathematical fit with historical temperaments also, is obtained with similar weights criteria; ($P_{MT}=0$; $P_{mt}=0.07$) for the collection of (K/N) marked WT, and ($P_{MT}=0$; $P_{mt}=0.0053$) for the full WT collection; all with $P_{MT}=0$.\\
The weighted Fifths/Thirds impurity ratios of the best ranked historical WT have values close to -1- (one), the ratio chosen for elaboration of the Beat-WT model. The fifths impurities show a slow and minor decrease, and have an almost parallel course with the weighted Fifths/Thirds impurity ratio.\\
The already mentioned downwards shift of Neidhardt and Werckmeister in table 9 is probably due to observable small deviations of thirds characteristics, see fig. \ref{fig. 10} and
 \ref{fig. 11}; probably also for Billeter, not here displayed. A remarkable symmetry can be observed, and the symmetry is perfect for the Beat-WT model. Symmetry is discussed in more detail in appendix B.2.

%

%
%
%
\section{Course of Diatonic Impurity, versus Tonality}
The C-major optimisation raises the question of what happens to other tonalities. It is to be expected that the unweighted diatonic impurity will increase with remote tonalities\footnote{Remote tonalities: tonalities that are not close to C-major, some are not admitted. See next footnote.}. Other question is how this stands comparing temperament families.

Table \ref{table 10} displays the numerical courses of the diatonic impurities of the 12TET, the meantone, Kirnberger III, and the unexplored EF-Beat-WT (§ 2.2.1.3; Appendix A.3 table \ref{table A1}).
Major tonalities that are admitted\footnote{Admitted tonalities: B$\flat$, F, C, G, D, A major, and corresponding minor keys G, D, A, E, B, F$\sharp$ minor} in meantone are displayed in the upper part of the table.

\begin{figure} [h]
\begin{center}
\includegraphics{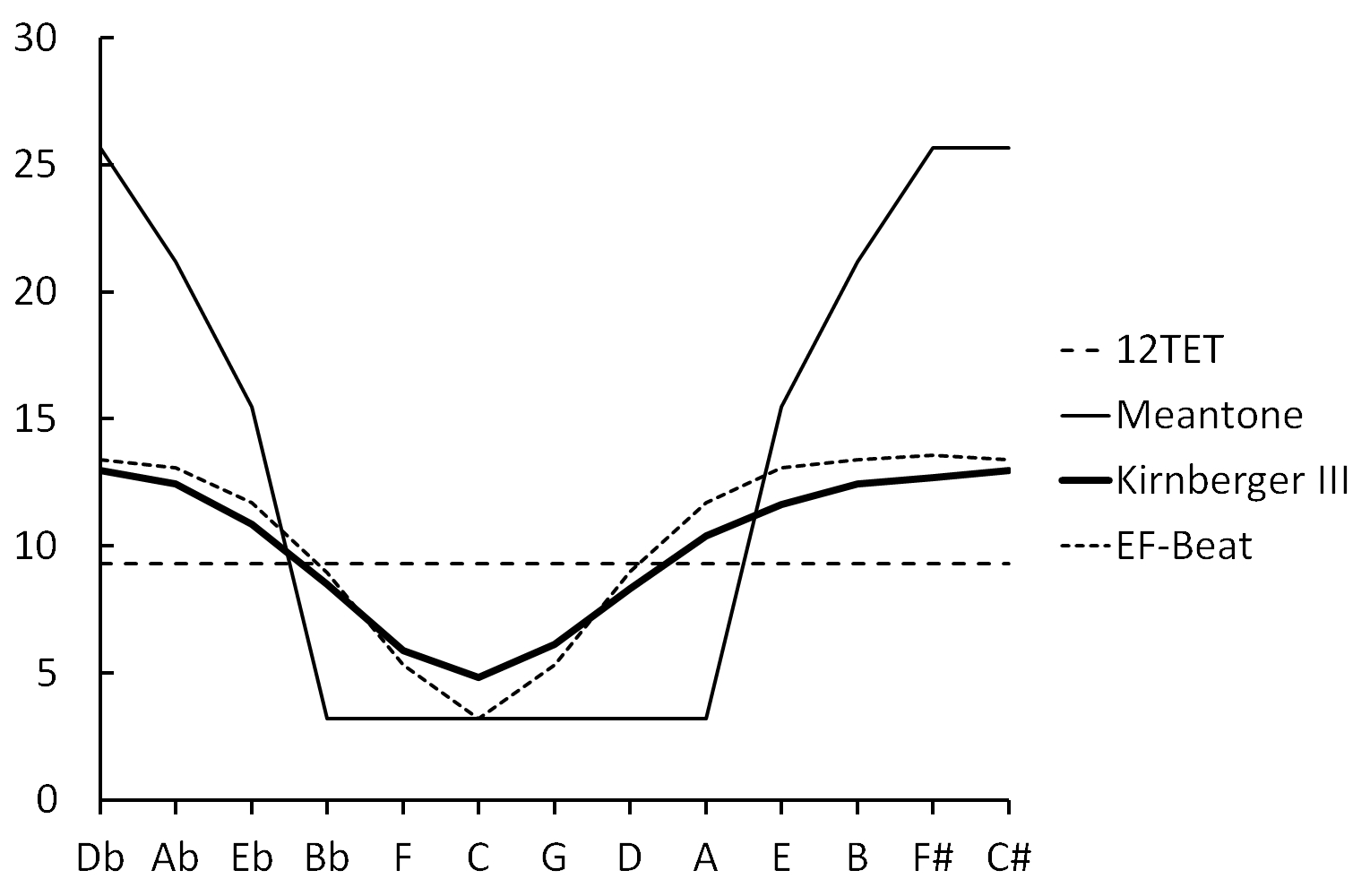}
\caption{Diatonic Impurity of Tonalities}
\label{fig. 12} 
\end{center}
\end{figure}

%
%
\begin{table} [h]
\begin{tabular}{|l|c|c|c|c|c|c|c|}
\hline
Admitted Tonalities & B$\flat$ & F & C & G & D & A \\
\hline
12TET & 9.30 & 9.30 & 9.30 & 9.30 & 9.30 & 9.30 \\
\hline
Meantone & 3.18 & 3.18 & 3.18 & 3.18 & 3.18 & 3.18 \\
\hline
Kirnberger III & 8.50 & 5.86 & 4.81 & 6.11 & 8.33 & 10.38 \\
\hline
EF-Beat-WT & 8.96 & 5.31 & 3.18 & 5.31 & 8.97 & 11.71 \\
\hline
\hline
Remote Ronalities & E & B & F$\sharp$ & C$\sharp$ / D$\flat$ & Ab & E$\flat$ \\
\hline
12TET & 9.30 & 9.30 & 9.30 & 9.30 & 9.30 & 9.30 \\
\hline
Meantone & 15.48 & 21.19 & 25.65 & 25.65 & 21.19 & 15.48 \\
\hline
Kirnberger III & 11.63 & 12.43 & 12.69 & 12.98 & 12.44 & 10.86 \\
\hline
EF-Beat-WT & 13.09 & 13.38 & 13.55 & 13.38 & 13.09 & 11.71 \\
\hline
\end{tabular}
\caption{Diatonic Beat Impurity of Tonalities}
\label{table 10}
\end{table}
The meantone has remarkable characteristics, and is of special interest still, for specific Baroque music: especially fig. \ref{fig. 12} dramatically illustrates the meantone characteristics course.

%

\begin{figure} [h]
\begin{center}
\includegraphics{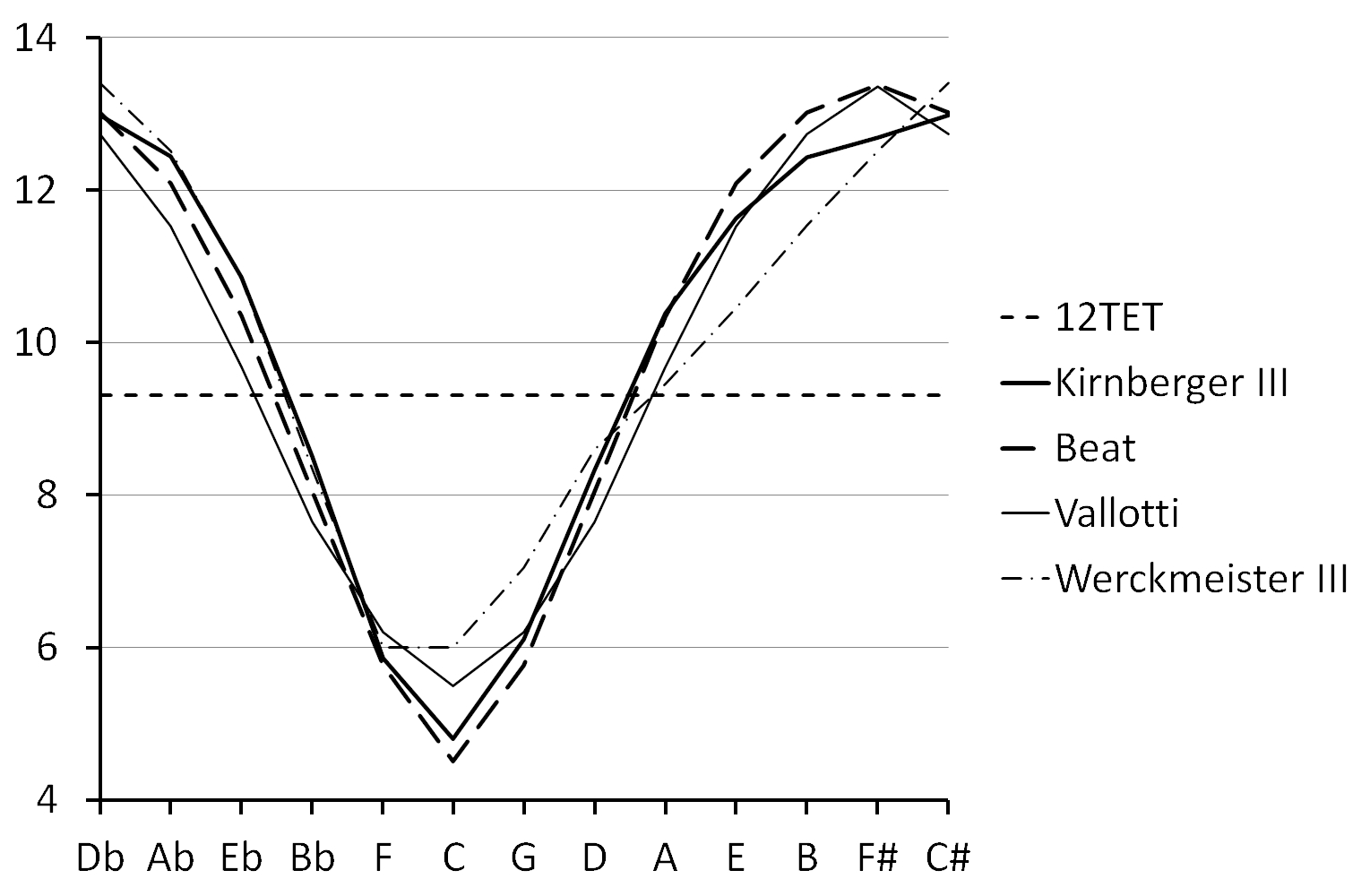}
\caption{Diatonic Impurity of WT Tonalities}
\label{fig. 13} 
\end{center}
\end{figure}

\begin{itemize}
\item \textbf{\emph{Surprisingly equal and excellent quality}} for all admitted tonalities, offering good modulations in a limited range of tonalities. Mozart, for example, usually stays within this domain (about 85\% of his corpus), exceptionally using the E or E$\flat$ key, but never as far as the B key \citep[][p. 11]{Kelletat1982}.
\item  \textbf{\emph{Excessive impurity, up to dissonance}}, for the remote tonalities.
\item \textbf{\emph{Specific meantone characteristics are sometimes deliberately exploited}}. A typical example consists of certain dramatic and highly dissonant passages of J. S. Bach's ``St. Matthew Passion''; for instance: proposal for crucifixion, death of Jesus, and a number of other outstanding dramatic events, \dots \citep[][p. 20]{Kelletat1982}.
Kelletat also cites Händel within this context, and many other examples can certainly be given by music connoisseurs and historians.
\end{itemize}
It is not always obvious to choose between meantone or a WT.
Possibilities for an artistic choice will remain, even within one family. A WT can be chosen among typical ones like those displayed in fig. \ref{fig. 10}, \ref{fig. 11}, and \ref{fig. 13}, or some with a flatter course like those below Werckmeister III in the ranking table \ref{table 8} or \ref{table 9}, or down to the perfectly flat 12TET as an extreme.

%

%
%
%
%

\section{Conclusions}
A number of models with ultimate diatonic C-major purity were developed.\\
Those models, were used for a proposed objective ranking of historical WT temperaments.
Widely prevailing musical appreciations of historical WT temperaments, and their obtained objective mathematical ranking, are in good agreement.\\
There is always a close fit of some selected better models with a famous and recognised historical temperament. Kirnberger III is always on top of the rankings, Werckmeister III is probably acceptable as a limit for recommendable WT.
\begin{center}
\textbf{\emph{Werckmeister's / Kelletat's musical WT definition}} \\
\textbf{\emph{ has gained an analysed mathematical support}} \\
\end{center}
The list of WT temperaments is very well filled, in small steps, starting at the models on top, down to the lower WT limit; -the 12TET temperament-.
\begin{center}
\textbf{\emph{The development of new WT achieves very little}} \\
\textbf{\emph{Because of artistic criteria, a need for choice of WT temperament persists.}} \\
\textbf{\emph{Even a need to leave the WT domain persists, and will not vanish.}} \\
\end{center}
\section*{Supplemental online material}
Supplemental online material for this article can be accessed at \url{doi-provided-by-publisher}.
It offers all calculations and solutions discussed in this paper, on a spreadsheet, and allows for observation of the behavior of musical intervals, in dependence of the entered weighting parameters.
\section*{Acknowledgements}
\textbf{\emph{Dedication}}\\

\noindent This paper is dedicated to all classical musicians and aural tuners, as homage to their musicality and refined musical ears.\\
\\
 \textbf{\emph{Acknowledgments}}\\
 
\noindent \textbf{\emph{A very special thanks is required here, namely to prof. E. Amiot.}} \\
He detected that one of my unpublished preceding texts, only based on results of iterations, was meaningful. Before his comments and suggestions, no weighting of intervals was applied, and no single attempt at analytical solution had been worked out. He has very strongly encouraged, promoted, and supported introduction, development, and application of solid mathematical methods, for acceptable presentation to scholars and interested academic community.\\
\textbf{\emph{It must with certainty be admitted that this analytical approach reveals remarkably much more insights on the subject, than the original primitive iterations.\\
In the mind of the author of this paper, the mathematical content belongs to prof. E. Amiot}}.\\

\noindent Thanks to my daughter Hilde: she advised me to investigate\quad \textbf{\emph{``what do musicians want''}}?,\quad instead of\quad ``what are the characteristics of some musician's favourite temperament''?\\
\noindent Thanks to my wife's friend, Margaret Roberts, neighbour and British citizen, for linguistic checks and improvements.
\\
\section*{Disclosure statement}
\addcontentsline{toc}{section}{Disclosure statement}

No potential conflict of interest was reported by the author. \\ \\ 

%
%

%
%
\appendices
%

%
\section{\qquad Formulas}
%
%
\subsection{ Formulas for\ § 2.2.1		First Model }
With: A=440\\

\noindent $\frac{\partial}{\partial C} \mathrm{:} \quad (25P_{MT}P_C^2+100P_{mt}P_A^2+9P_C^2+ 16P_F^2)C -20P_{MT}P_C^2E-12P_{F}^2F-6P_C^2G$\\

\qquad $=60P_{mt}P_A^2A$\\

\noindent $\frac{\partial}{\partial D}\mathrm{:}\quad (36P_{mt} P_D^2+100P_{mt}P_B^2+9 P_D^2+16P_G^2 )D-30P_{mt} P_D^2 F -12P_G^2 G  -60P_{mt} P_B^2 B$\\

\qquad $=6 P_D^2 A$\\

\noindent $\frac{\partial}{\partial E}\mathrm{:}\quad -20P_{MT} P_C^2 C+(16P_{MT} P_C^2+36P_{mt} P_F^2+9 P_E^2+16 P_A^2 )E-30P_{mt} P_E^2 G-6 P_E^2 B$\\

\qquad $=12 P_A^2 A$\\

\noindent $\frac{\partial}{\partial F}\mathrm{:}\quad -12 P_F^2 C-30P_{mt} P_D^2 D+(25P_{MT} P_F^2+25P_{mt} P_D^2+9 P_F^2 )F=20P_{MT} P_E^2 A$\\

\noindent $\frac{\partial}{\partial G}\mathrm{:}\quad -6 P_C^2 C-12 P_G^2 D-30P_{mt} P_E^2 E +(25P_{MT} P_G^2+25P_{mt} P_E^2+4 P_C^2+9 P_G^2 )G$\\

\qquad $-20P_{MT} P_G^2 B=0$ \\

\noindent $\frac{\partial}{\partial B}\mathrm{:}\quad -60P_{mt} P_B^2 D-6 P_E^2 E-20P_{MT} P_G^2 G+(16P_{MT} P_G^2+36P_{mt} P_B^2+4 P_E^2 )B=0$\\

\noindent The obtained $B$ and $F$ values have to be entered in equations concerning altered notes.\\

\noindent $\frac{\partial}{\partial C \sharp}\mathrm{:}\quad (9P_{C\sharp}^2+16P_{F\sharp}^2 )C\sharp-12P_{F\sharp}^2 F\sharp-6P_{C\sharp}^2 G\sharp=0 $\\

\noindent $\frac{\partial}{\partial F\sharp}\mathrm{:}\quad -12P_{F\sharp}^2 C\sharp+(9P_{F\sharp}^2+16P_B^2 )F\sharp=12P_B^2 B $\\

\noindent $\frac{\partial}{\partial G\sharp}\mathrm{:}\quad -6P_{C\sharp}^2 C\sharp+(4P_{C\sharp}^2+9P_{G\sharp}^2 )G\sharp-12P_{G\sharp}^2 E\flat=0 $\\

\noindent $\frac{\partial}{\partial E\flat}\mathrm{:}\quad -12P_{G\sharp}^2 G\sharp+(16P_{G\sharp}^2+9P_{E\flat}^2 )E\flat-6P_{Eb}^2 B\flat=0 $\\

\noindent $\frac{\partial}{\partial B\flat}\mathrm{:}\quad -6P_{E\flat}^2 E\flat+(4P_{E\flat}^2+9P_{B\flat}^2 )B\flat=12P_{B\flat}^2 F $\\

%
%
\subsection{ Formulas for\ § 2.2.2\quad		Domain for $B=kF$, with $k=2^{10}/3^{6}$}
With: A=440 and $k=2^{10}/3^6$ \\

\noindent $\frac{\partial}{\partial C}\mathrm{:}\quad (25P_C^2 P_{MT}+100P_A^2 P_{mt}+9P_C^2 +16P_F^2  )C -20P_C^2 P_{MT} E-12P_F^2  F-6P_C^2  G$\\

\qquad $=60P_A^2 P_{mt} A $\\

\noindent $\frac{\partial}{\partial D}\mathrm{:}\quad (36P_D^2 P_{mt}+100P_B^2 P_{mt}+9P_D^2 +16P_G^2  )D -(30P_D^2 P_{mt}+60kP_B^2 P_{mt} )F -12P_G^2  G$\\

\qquad $=6P_D^2  A$ \\

\noindent $\frac{\partial}{\partial E}\mathrm{:}\quad -20P_C^2 P_{MT} C+(16P_C^2 P_{MT}+36P_E^2 P_{mt}+9P_E^2 +16P_A^2  )E-6kP_E^2  F  -30P_E^2 P_{mt} G$\\

\qquad $=12P_A^2  A $ \\

 \noindent $\frac{\partial}{\partial F}\mathrm{:}\quad +(25P_F^2 P_{MT}+16k^2 P_G^2 P_{MT}+25P_D^2 P_{mt}+36k^2 P_B^2 P_{mt}+4k^2 P_E^2 +9P_F^2  )F-12P_F^2  C$\\
 
\qquad $ -(30P_D^2 P_{mt}+60kP_B^2 P_{mt} )D-6kP_E^2  E-20kP_G^2 P_{MT} G =20P_F^2 P_{MT} A$ \\

\noindent $\frac{\partial}{\partial G}\mathrm{:}\quad -6P_C^2  C-12P_G^2  D-30P_E^2 P_{mt} E-20kP_G^2 P_{MT} F$\\

\qquad $+(25P_G^2 P_{MT}+25P_E^2 P_{mt} +4P_C^2 +9P_G^2 )G=0 $ 
%
%
\subsection{Solution for\ § 2.2.1.3 \quad		Domain above the AB line, fig. \ref{fig. 5} }

The model of § 2.2.1, for $P_{MT}=P_{mt}=1$, has major thirds on E, B, F$\sharp$, C$\sharp$ and G$\sharp$ exceeding the 81/64 ratio, and major thirds on F, C and G below the 5/4 ratio. This goes against the formula 2 conditions. Compliance with those conditions was achieved in § 2.2.2, by presetting the AB-line condition (see § 2.2.1.1, setting B=kF, and fig. 5).

An alternate way to achieve compliance with formula 2, might be, in a first step, to preset an 81/64 ratio for thirds standing on the diatonic major thirds (e.g. E-G$\sharp$ and G$\sharp$-C, on C-E), see fig. A1. This involves all above mentioned major thirds, except the one on F$\sharp$.

\begin{figure} [h]
\begin{center}
\includegraphics{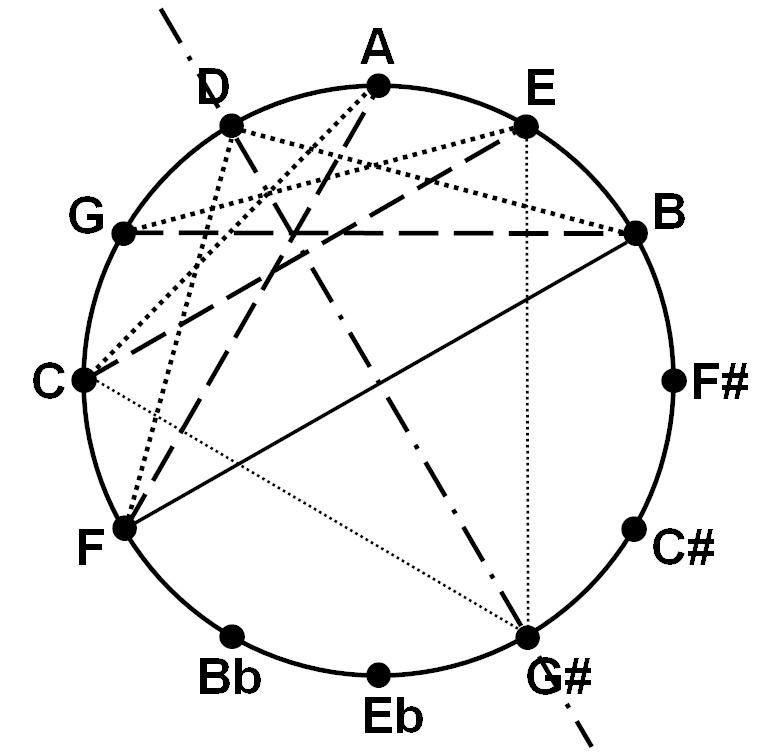}
\caption{Important Diatonic Intervals}
\label{fig. A1} 
\end{center}
\end{figure}

The obtained models have up to seven major thirds with the 81/64 ratio, but none exceeds this ratio; the major thirds on F, C and G have a ratio $m=3^8/2^{13}$, still below the 5/4 ratio; the models are therefore, still not complying. The model interval courses are similar to those displayed in fig. A2. Those courses reveal that the required just major thirds on F, C and G can be obtained, by a suitable adaptation of involved thirds:
\begin{itemize}
\item Maintain a pythagorean major third on B and C$\sharp$, with a just third on F and G; this also leads to complying fixed 512/405 ratio thirds on E$\flat$ and A
\item The just third on C: leave the thirds on E and G$\sharp$ unknown (hence: G$\sharp$=variable);
\end{itemize}

\noindent With m=4/5, the above leads to following substitutions in formulas 13 and 14:\\

\quad	$F=mA$;\quad $C=mE$;\quad $G=mB$;$\quad C\#={\frac{64}{81}}F$;\quad $Eb={\frac{81}{128}} B $ \\
	
\noindent Partial derivatives for some notes expire, and so do the expressions for fixed major thirds. The remaining partial derivatives of formulas 13 and 14, set to zero, lead to:\\

\noindent $\frac{\partial}{\partial D}\mathrm{:}\quad (36P_D^2 P_{mt}+100P_B^2 P_{mt}+9P_D^2 +16P_G^2  )D-(60P_B^2P_{mt}+12mP_G^2  )B $\\

\qquad $ =(30mP_D^2 P_{mt}+6P_D^2  )A $ \\

\noindent $\frac{\partial}{\partial E}\mathrm{:}\quad (36P_E^2 P_{mt}+100m^2 P_A^2 P_{mt}+9m^2 P_C^2 +9P_E^2 +16m^2 P_F^2 +16P_A^2  )E $\\

\qquad $ -(30mP_E^2 P_{mt}+6m^2 P_C^2 +6P_E^2  )B=(60mP_A^2 P_{mt}+12m^2 P_F^2 +12P_A^2  )A $ \\

\noindent $\frac{\partial}{\partial B}\mathrm{:}\quad -(60P_B^2 P_{mt}+12mP_G^2  )D-(30mP_E^2 P_{mt}+6m^2 P_C^2 +6P_E^2  )E $\\

\qquad $+(25m^2 P_E^2 P_{mt}+36P_B^2 P_{mt}+4m^2 P_C^2 +4P_E^2 +9m^2 P_G^2  )B=0 $ \\

\noindent Only F$\sharp$,  G$\sharp$ and B$\flat$ remain to be solved, based on the obtained B and F values.\\

\noindent $\frac{\partial}{\partial F\sharp}\mathrm{:}\quad (16P_B^2+9P_{F\sharp}^2 )F\sharp={\frac{256}{27}} P_{F\sharp}^2 F+12 P_B^2 B $\\
 
\noindent $\frac{\partial}{\partial G\sharp}\mathrm{:}\quad (4P_{C\sharp}^2+9P_{G\sharp}^2)G\sharp=\frac{243}{32}P_{C\sharp}^2F+\frac{243}{32}P_{G\sharp}^2B $\\

\noindent $\frac{\partial}{\partial B\flat}\mathrm{:}\quad (4P_{E\flat}^2+9P_{B\flat}^2 )B\flat=12 P_{B\flat}^2 F + {\frac{243}{64}} P_{E\flat}^2 B$ \\

\begin{figure} [h]
\begin{center}
\includegraphics{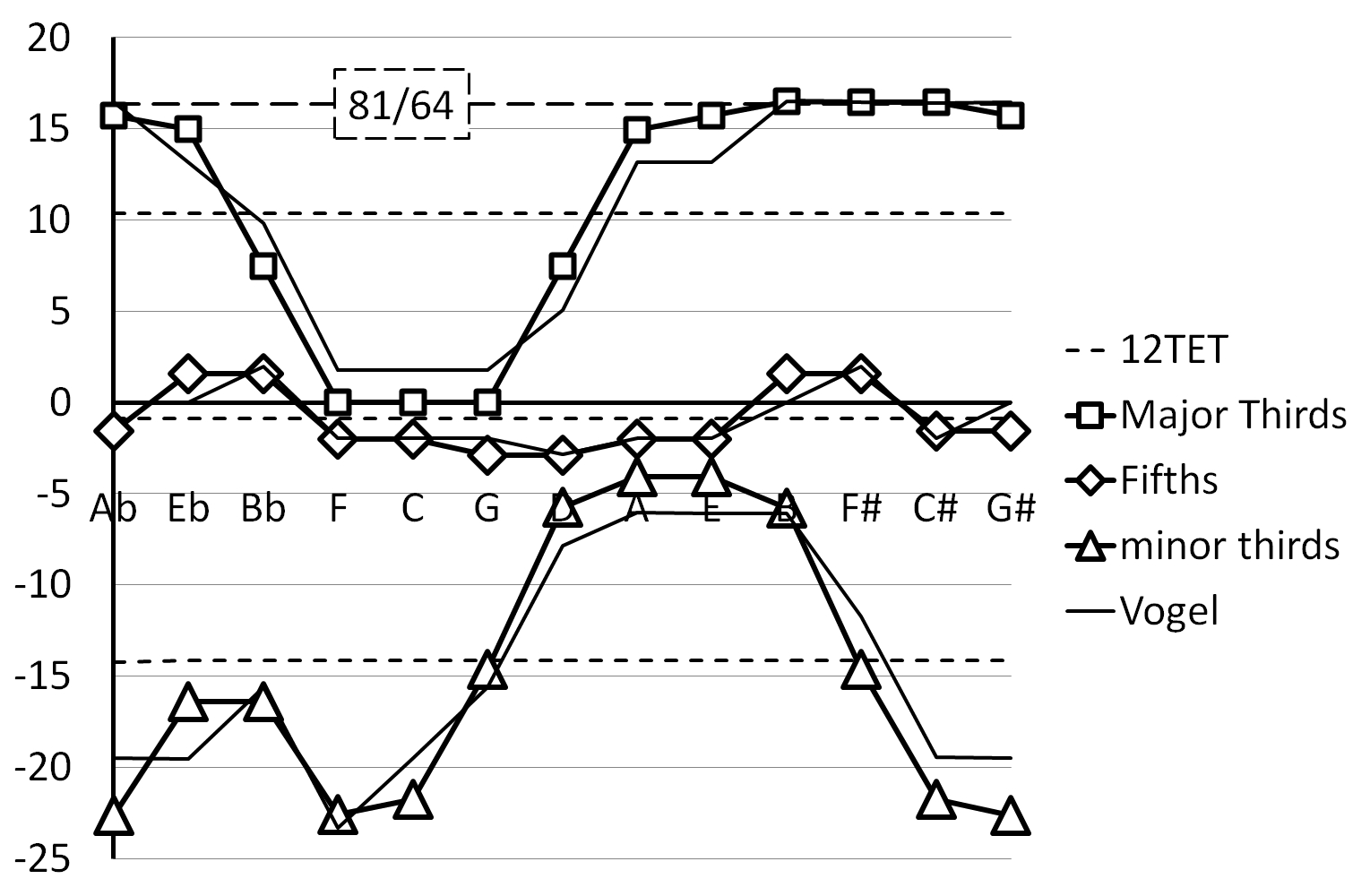}
\caption{Ultimate WT: EF-Beat-WT}
\label{fig. A2} 
\end{center}
\end{figure}

\noindent The above conditions, combined with the condition of balanced thirds and fifths (see § 2.2.2; formula 17, and fig.  \ref{fig. 9}), lead with (PMT=0; Pmt=.33287) to the model holding pitches displayed in table A1:
%
%
\begin{table} [h]
\begin{tabular}{|l|c|c|c|c|c|c|}
\hline
\textbf{\emph{EF-Beat-WT}} & C & C$\sharp$ & D & E$\flat$ & E & F \\
\hline
Pitches & 263.32 & 278.12 & 294.40 & 311.64 & 329.15 & 352.00\\
\hline
\hline
\textbf{\emph{EF-Beat-WT}} &  F$\sharp$ & G & G$\sharp$ & A & B$\flat$ & B \\
\hline 
Pitches & 370,09 & 393.97 & 416.35 & 440.00 & 468.39 & 492.47\\
\hline
\end{tabular}
\caption{EF-Beat-WT model  \qquad 	(*) EF: Enlarged Fifth}
\label{table A1}
\end{table}\\
Fig. \ref{fig. A2} displays the interval characteristics of this model, also in comparison of those of Vogel (see also § 3.1.1, table \ref{table 8}, and comments). This model's unweighted diatonic purity is better than Vogel's one: 3.184 against 3.997.\\
\\
\textbf{\emph{Musical differences with historical WT are prominent (see fig. \ref{fig. A2})}}.
%
%
\begin{itemize}
\item Pythagorean major thirds: This model contains seven major thirds, very close or equal to the 81/64 Pythagorean ratio.
\item Fifths containing an altered note: four of those fifths are enlarged and grouped in a specific way, allowing persistent fifths enlargement, without inducing major thirds to exceed the 81/64 Pythagorean ratio. 
\end{itemize}
\textbf{\emph{This model and Vogel are part of a possible but yet unexplored musical WT domain}}.\\

\noindent Musical practice based on this model might be topic of musical experimentation, this falls however outside the scope of this paper.
%
%

\section{\qquad Application of Interval RATIO Measurements}
%
\subsection{Relation between interval ratio, and interval impurity measurement}
Interval ratio impurity measurements, like PBP, PPD or Cent (formula \ref{equ 7}-\ref{equ 9}), allow for simplifications.\\
\begin{equation}
\mathrm{Let \quad} \frac{x_2}{x_1} \mathrm{, }\frac{x_3}{x_2}\mathrm{ ,\dots ,}\frac{x_n}{x_{n-1}} \mathrm{ , \dots  ,}\frac{kx_1}{x_p} \mathrm{ \quad for\quad n=1 \ to\ p;\ and \quad k=\quad  a\ fixed\ value}\\
\label{equ B1}
\end{equation}
\noindent be a series of intervals (ratios), wherefore the expression below should be minimised.\\
\begin{equation}
F_x(x_1 \mathrm{, \dots} x_n \mathrm{, \dots}x_p) = \left [F_a \left (\frac{x_2}{x_1} \right ) \right ]^2+ \left  [F_a \left (\frac{x_3}{x_2} \right  ) \right  ]^2 \mathrm{ \dots}+ \left [F_a \left (\frac{x_n}{x_{n-1}} \right  ) \right  ]^2 \mathrm{ \dots}+ \left [F_a \left (\frac{kx_1}{x_p} \right ) \right  ]^2
\label{equ B2}
\end{equation}
Setting the partial derivatives to zero, this is        $\frac{\partial F_x}{\partial x_n}=0$     for $n=1$ to $p$, leads to
\begin{equation}
F_a \left (\frac{x_n}{x_{n-1}} \right  )=F_a \left (\frac{x_{n+1}}{x_n} \right  ) \mathrm{\quad for \ }n=2 \mathrm{\quad to\quad } (p-1) \mathrm{\quad but\ also\quad }
\label{equ B3}
\end{equation}
\begin{equation}
F_a \left (\frac{x_2}{x_1} \right  )=F_a \left (\frac{kx_1}{x_p} \right  ) \mathrm{\quad for \ }n=1 \mathrm{\quad and \quad } F_a \left (\frac{x_p}{x_{p-1}} \right  )=F_a \left (\frac{kx_1}{x_p}  \right )  \mathrm{\quad for \ }n=p
\label{equ B4}
\end{equation}
Hence, all functions	$F_a \left (\frac{x_{n+1}}{x_n } \right )$  for $n=1$ to  $(p-1)$, and $F_a \left (\frac{kx_1}{x_p} \right )$ must have equal value.\\

A possible solution, not excluding any other one, therefore contains the equalities
\begin{equation}
\frac{x_2}{x_1}=\frac{x_3}{x_2}=\mathrm{ ,\dots,}=\frac{x_n}{x_{n-1}}= \mathrm{ ,\dots,}\frac{x_p}{x_{p-1}}=\frac{kx_1}{x_p}=C_t \\
\label{equ B5}
\end{equation}
\textbf{\emph{Conclusion:\\
For a group composed of equal type intervals, the best group quality is obtained at equality of ratio of all intervals, if the impurity measurement is based on measuring the ratios of the note pitches of the intervals.}}
%
%
\subsection{Models Symmetry}
Examination of optimised Beat models revealed remarkable symmetrical characteristics, see fig. \ref{fig. 11}, \ref{fig. 12}, and fig. \ref{fig. A2}. Further examination of the circle of fifths fig. \ref{fig. A1} reveals furthermore:

\begin{itemize}
\item A symmetric geometric structure can be observed on the circle of fifths, for intervals concerned in the calculation of the diatonic impurity: fig. \ref{fig. A1} depicts fifths as thirty degree arcs on the circle, thirds as arc cords, drawn as striped lines (major thirds) or dotted lines (minor thirds).\\ A symmetry axis passes by the notes D and G$\sharp$.\\ Notice: on the keyboard too, D and G$\sharp$ are (optical) symmetry centers
\item Because of, and thanks to this symmetry \textbf{\emph{and ratio measuring units}}, we have :
\begin{enumerate}
\item equality of diatonic fifths F-C and E-B; C-G and A-E; G-D and D-A
\item equality of diatonic major thirds F-A and G-B
\item equality of diatonic minor thirds D-F and B-D; A-C and E-G
\item equality of fifths with altered notes: B-F$\sharp$ and B$\flat$-F; F$\sharp$-C$\sharp$ and E$\flat$-B$\flat$; C$\sharp$-G$\sharp$ and G$\sharp$-E$\flat$ ($\div$A$\flat$-E$\flat$)
\item D and G$\sharp$ are exactly diametrically opposite
\item a reduction of the number of variable notes: reduction to three for the C-major diatonic notes, and two for the altered notes.\\ 
The chosen notes have to be paired and independent\footnote{Most notes come in pairs, because of symmetry. If a pair, then one of both should be selected.}, we therefore have:
\begin{itemize}
\item For the C-major diatonic notes: F (or B), C (or E) and D
\item For the altered notes: C$\sharp$ (or E$\flat$) and F$\sharp$ (or B$\flat$); (G$\sharp$ depends on D: point 5)
\end{itemize}
\end{enumerate}
\end{itemize}
%
%
\subsubsection{Substitutions / possible eliminations}
Some notes can be substituted, in line with the observations at Section B.2.
\begin{equation}
\mathrm{}
\mathrm{Notes\ C\ or\ E: \quad we\ have\quad}	\frac{E}{D}=\frac{D}{C}       \mathrm{\quad hence\quad}   C=\frac{D^2}{E}
\label{equ B6}
\end{equation}
\begin{equation}
\mathrm{}
\mathrm{Note\ G:\quad we\ have,\ counting\ the\ octaves\quad}	\frac{A}{D}=\frac{2D}{G}       \mathrm{\quad hence\quad}   G=\frac{2D^2}{AE}
\label{equ B7}
\end{equation}
Notes F or B; we have, counting the octaves
\begin{equation}
\frac{B}{D}=\frac{2D}{F} \qquad       \mathrm{\ hence} \qquad   F=\frac{2D^2}{B}
\label{equ B8}
\end{equation}
Notes D and G$\sharp$ being diametrically opposite, we have
\begin{equation}
 \qquad      G\sharp=D\sqrt{2}
 \label{equ B9} 
\end{equation}\\
%
%
\paragraph{Possible additional substitution (for models with $B/F =2^{10}/3^6$)}
Six perfect fifths between B to F, containing an altered note, means also three perfect fifths from G$\sharp$ to F. Setting F and B to the same octave, it can be said.
\begin{equation}
B= \frac{2^{10}}{3^6} F \mathrm{\quad and\quad }	F=\frac{3^3}{2^5 }G\sharp=\frac{3^3}{2^5 }\sqrt{2}D \mathrm{\quad hence\quad }	B=\frac{2^{5.5}}{3^3} D
\label{equ B10}
\end{equation}
%
%
\paragraph{Possible additional elimination (for WT models with $B/F =2^{10}/3^6$)}
See fig. \ref{fig. A1}. The line connecting F to B stands perpendicular to the G$\sharp$-D diameter and has a fixed value, because $B/F=2^{10}/3^6$; the lines F-D and D-B therefore have a fixed value too. In other words: \textbf{\emph{the minor thirds B-D et D-F have a fixed value, totally independent of the $P_{MT}$ and $P_{mt}$ weighing parameters}}; their ratio equals $2^5/3^3=1.185…$\\
The expressions representing \textbf{\emph{those minor thirds impurity may consequently be eliminated in the sum of squares}}, formula \ref{equ 13}, while, being constants, their partial derivatives will be zero.
%
%
\subsubsection{Model calculation with Cents (for WT models with $B/F=2^{10}/3^6$)}
Impurity formulas: expressions for minor thirds impurities on notes D and B can be eliminated (see § B.2.1.2).

%
\begin{table} [h]
\begin{tabular}{|l|l|}
\hline
Major thirds impurities: & Fifths impurities \\
C: $\Delta MT_C=1200\log{(4E/5C)}/\log{2}$ & C: $\Delta Fi_C=1200\log{(2G/3C)}/\log{2}$ \\
F: $\Delta MT_F=1200\log{(4A/5F)}/\log{2}$ & D: $\Delta Fi_D=1200\log{(2A/3D)}/\log{2}$ \\
G: $\Delta MT_G=1200\log{(4B/5G)}/\log{2}$ & E: $\Delta Fi_E=1200\log{(2B/3E)}/\log{2}$ \\
minor thirds impurities: & F: $\Delta Fi_F=1200\log{(4C/3F)}/\log{2}$ \\
D: eliminated & G: $\Delta Fi_G=1200\log{(4D/3G)}/\log{2}$ \\
E: $\Delta mt_E=1200\log{(5G/6E)}/\log{2}$ & A: $\Delta Fi_A=1200\log{(4E/3A)}/\log{2}$ \\
A: $\Delta mt_A=1200\log{(10C/6A)}/\log{2}$ &  B: A diatonic fifth on B does not exist: \\
B: eliminated & a perfect fifth on B leads indeed to F\# \\
\hline
\end{tabular}
\caption{Diatonic C-major impurities}
\label{table B1}
\end{table}
\noindent The notes F, C, G, and B can be substituted: formulas (\ref{equ B6}; \ref{equ B7}; \ref{equ B8}; \ref{equ B10}).\\
With $k=3^3/2^{4,5}$  the sum of squares of impurities becomes
%
\begin{displaymath}
P_{MT} \left \{ \left [ \log {\left (\frac{4E^2}{5D^2} \right )} \right ]^2 + 2 \left [ \log {\left (\frac{4A}{5kD} \right )} \right ]^2  \right \}+ 2P_{mt} \left [ \log {\left (\frac{5D^2}{3AE} \right )} \right ]^2      
\end{displaymath}
\begin{equation}
+2\left \{ \left [ \log {\left (\frac{4E}{3A} \right )} \right ]^2 + \left [ \log {\left (\frac{2A}{3D} \right )} \right ]^2  +   \left [ \log {\left (\frac{4D}{3kE} \right )} \right ]^2 \right \}
\label{equ B11}
\end{equation}
The partial derivatives of remaining variables D and E, set to zero, lead to\\

%
\noindent $\frac{\partial}{\partial D}\mathrm{:}\quad   (3P_{MT}+4P_{mt}+2 )\log{D}   -   (2P_{MT}+2P_{mt}+1)\log{E} $\\

\qquad $ =(4P_{MT}-1)\log2 +2P_{mt}\log3-2(P_{MT}+P_{mt})\log5  -(P_{MT}-1)\log{k}$
\begin{equation}
+(P_{MT}+2P_{mt}+1)\log{A} \qquad \qquad \qquad \qquad \qquad \qquad \qquad  \qquad
\label{equ B12}
\end{equation}
\noindent $\frac{\partial}{\partial E}\mathrm{:}\quad  - (2P_{MT}+2P_{mt}+1 )\log{D}   +   (2P_{MT}+P_{mt}+2)\log{E} $
\begin{equation}
 =-2P_{MT}\log2-P_{mt}\log3+(P_{MT}+P_{mt})\log5 -P_{Fi}\log{k} -(P_{MT}-1)\log{A}
 \label{equ B13}
 \end{equation}
The solution can be calculated in a spreadsheet that can be downloaded.\\
Pitches of Kirnberger III, and those obtained here for this Cent-WT model with balanced impurity of thirds and fifths (applying formula \ref{equ 17}, see also fig. \ref{fig. 9}), and those of the Beat-WT model § 2.2.2, table \ref{table 7}, are very similar (see table \ref{table B2}):
%
%
\begin{table} [h]
\begin{tabular}{|l|c|c|c|c|c|c|}
\hline
& C & C$\sharp$ & D & E$\flat$ & E & F \\
\hline
\textbf{\emph{Cent-WT}} & 263.24 & 277.56 & 294.40 & 312.26 & 329.26 & 351.29\\
\hline
\textbf{\emph{Beat-WT}} & 263.23 & 277.55 & 294.39 & 312.25 & 329.25 & 351.28\\
\hline
\textbf{\emph{Kirnberger III}} & 263.20 & 277.32 & 294.20 & 311.94 & 329.00 & 350.93\\
\hline
\hline
 &  F$\sharp$ & G & G$\sharp$ & A & B$\flat$ & B \\
\hline
\textbf{\emph{Cent-WT}} & 370.08 & 393.96 & 416.35 & 440.00 & 468.39 & 493.45\\
\hline
\textbf{\emph{Beat-WT}} & 370.07 & 393.95 & 416.33 & 440.00 & 468.37 & 493.43\\
\hline 
\textbf{\emph{Kirnberger III}} & 370,13 & 393.50 & 415.98 & 440.00 & 467.91 & 493.50\\
\hline
\end{tabular}
\caption{Comparison of Kirnberger III, and optimal Cent and Beat models}
\label{table B2}
\end{table}
\subsection{Kirnberger III}
A preset of a just major third on C, taking into account all preceding considerations of this appendix, 
allows for following additional substitution of E in proportion to D, so that only one variable remains: the note D:
\begin{displaymath}
E=\frac{5C}{4}\quad \mathrm{but} \quad  C=\frac{D^2}{E} \quad \mathrm{(form.\ \ref{equ B6});\ hence} \quad E=\frac{\sqrt{5}}{2}D
\end{displaymath}
\subsubsection{Cent Model}
The derivative by D, of the sum of squares, set to zero (with $ k=3^3/2^{4.5}$), leads to
\begin{displaymath}
\log{D}=\log{A}+\frac{P_{MT} \log{(\frac{4}{5k})}-P_{mt} \log{(\frac{2 \sqrt{5}}{3})}- \log{(\frac{2 \sqrt{5}}{3})}+ \log{(\frac{2}{3})}}{P_{MT}+P_{mt}+2}
\end{displaymath}
Model with balanced thirds/fifths: $RMS$-$\Delta$-$Cent=1.18$ if compared to Kirnberger III.
\subsubsection{Beat Model}
Based on the same symmetry considerations as for the ratio based Cent and PBP models, a Beat model (Beat / PBP equivalence: see § 1.4.2) can be solved . The outcome (with $k=3^3/2^{4.5}$) leads to
\begin{displaymath}
D=\frac{20kP_{MT}P_F^2+\frac{120P_{mt}P_A^2}{\sqrt{5}}+6\sqrt{5}P_A^2+6P_D^2}{25kP_{MT}P_F^2+80P_{mt}P_A^2+20P_A^2+9P_D^2}A
\end{displaymath}
Model with balanced thirds/fifths: $RMS$-$\Delta$-$Cent=1.18$ if compared to Kirnberger III.\\
This model is almost identical to the above Cent model, § B.3.1 .
\subsubsection{Optimisation of fifths only (Cent and Beat)}
With $P_{MT}=P_{mt}=0$ the outcome is:
\begin{itemize}
\item Fifths on C, G, D, A: $ratio = 5^{(1/4)}$
\item Fifths on B, F$\sharp$, C$\sharp$, G$\sharp$, E$\flat$, B$\flat$: $ratio = 3/2$
\item Fifths on F, E: $ratio=(2^6/3^3)(2/5)^{(1/2)}$
\end{itemize}
Model comparison: $RMS$-$\Delta$-$Cent=0.945$ if compared to Kirnberger III.
\newpage
\noindent 
Fig 01.jpg: (fig. 1) Orgelprobe 1681\\
Fig 02.jpg: (fig. 2) C-major diatonic structure\\
Fig 03.jpg: (fig. 3) Beating of a musical interval\\
Fig 04.jpg: (fig. 4) Impact of squares\\
Fig 05.jpg: (fig. 5) Diatonic Impurity as $F(P_{MT}$; $P_{mt}$)\\
Fig 06.jpg: (fig. A1) Important Diatonic Intervals\\
Fig 07 new.jpg: (fig. 6) Diatonic Impurity as $F(P_{MT}$; $P_{mt}$)\\
Fig 08.jpg: (fig. 7) Course of unweighted diatonic thirds and fifths\\
Fig 09.jpg: (fig. 8) Course of weighted diatonic thirds and fifths\\
Fig 10.jpg: (fig. 9) Historical WT characteristics ranked according table \ref{table 9}\\
Fig 11.jpg: (fig. 10) Beat impurity of major thirds (up) and fifths (down)\\
Fig 12.jpg: (fig. 11) Beat impurity of minor thirds\\
Fig 13.jpg: (fig. 12) Diatonic Impurity of Tonalities\\
Fig 14.jpg: (fig. 13) Diatonic Impurity of WT Tonalities\\
Fig 15.jpg: (fig. A2) Ultimate WT: EF-Beat-WT\\
\\
Table 01: Note weights applicable with Beat impurity measurements\\
Table 02: Diatonic C-major impurities\\
Table 03: Impurities of fifths containing an altered note\\
Table 04: Weighted Diatonic Impurity\\
Table 05: Beat-WT-A and Beat-WT-B\\
Table 06: Impurities of Fifths and the C-major diatonic scale\\
Table 07: impurity = 2.7024; (4.2928 unweighted) Beat-WT $P_ {MT}=0$; $P_ {mt}=0.1177$\\
Table 08: WT Diatonic C-major Impurities\\
Table 09: $RMS$-$\Delta$-$Cents$ comparison of historical WT temperaments with Beat model\\
Table 10: Diatonic Beat Impurity of Tonalities\\
\\
Table A1: EF-Beat-WT model\quad  (*) EF: Enlarged Fifth\\
\\
Table B1: Diatonic C-major impurities\\
Table B2: Comparison of Kirnberger III, and optimal Cent and Beat models\\\

\end{document}